\documentclass[useAMS,usenatbib]{mn2e}
\usepackage{url}
\usepackage{graphicx}
\usepackage{lscape}
\usepackage{amsmath,amssymb}
\usepackage{longtable}
\usepackage{deluxetable}

\usepackage{color}

\usepackage{soul}   

\newcommand{\HI}{\hbox{\rmfamily H\,{\textsc i }}}

\newcommand{\HIsub}{\hbox{{\scriptsize H}\,{\tiny I}}}
\newcommand{\HII}{\hbox{\rmfamily H\,{\scshape ii}}}

\newcommand{\Ha}{\hbox{\rmfamily H\,$\alpha$}}
\newcommand{\msun}{\hbox{$M_{\odot}$}}

\title[The star formation relation for LV dwarf galaxies]{The disk averaged star formation relation for Local Volume dwarf galaxies}

\author[A. L\'{o}pez-S\'{a}nchez et al.]{\'{A}. R. L\'{o}pez-S\'{a}nchez$^{1,2,3}$\thanks{angel.lopez-sanchez@aao.gov.au}, C.D.P. Lagos$^{4,5,3}$, T. Young$^{2,6,7}$, H. Jerjen$^{6}$\\
$^{1}$ Australian Astronomical Observatory, 105 Delhi Road, North Ryde, NSW 2113, Australia \\
$^{2}$ Department of Physics and Astronomy, Macquarie University, NSW 2109, Australia\\
$^{3}$ Australian Research Council Centre of Excellence for All Sky Astrophysics in 3 Dimensions (ASTRO 3D), Australia\\
$^{4}$ International Centre for Radio Astronomy Research (ICRAR), M468, University of Western Australia, 35 Stirling Hwy, \\Crawley, WA 6009, Australia.\\
$^{5}$ Australian Research Council Centre of Excellence for All-sky Astrophysics (CAASTRO), 44 Rosehill Street\\ Redfern, NSW 2016, Australia.\\
$^{6}$ Research School of Astronomy and Astrophysics, The Australian National University, Mt Stromlo Observatory, via Cotter Rd,\\ Weston, ACT 2611, Australia\\
$^{7}$ CSIRO Astronomy and Space Science, Australia Telescope National Facility, PO Box 76, Epping, NSW 1710, Australia}

\begin{document}

\date{}

\pagerange{\pageref{firstpage}--\pageref{lastpage}} \pubyear{}

\maketitle

\label{firstpage}

\begin{abstract}
Spatially resolved \HI studies of dwarf galaxies have provided a wealth of precision data. However these high-quality, resolved observations are only possible for handful of dwarf galaxies in the Local Volume. Future \HI surveys are unlikely to improve the current situation. We therefore explore a method for estimating the surface density of the atomic gas from global \HI parameters, which are conversely widely available. We perform empirical tests using galaxies with resolved \HI maps, and find that our approximation produces values for the surface density of atomic hydrogen within typically \mbox{0.5 dex} of the true value. We apply this method to a sample of $147$ galaxies drawn from modern near-infrared stellar photometric surveys. With this sample we confirm a strict correlation between the atomic gas surface density and the star formation rate surface density, that is vertically offset from the Kennicutt-Schmidt relation by a factor of  $10-30$, and  significantly steeper than the classical $N=1.4$ of Kennicutt (1998). We further infer the molecular fraction in the sample of low surface brightness, predominantly dwarf galaxies by assuming that the star formation relationship with molecular gas observed for spiral galaxies also holds in these galaxies, finding a molecular-to-atomic gas mass fraction within the range of 5-15\%. Comparison of the data to available models shows that a model in which the thermal pressure balances the vertical gravitational field captures better the shape of the $\Sigma_\text{SFR}$-$\Sigma_\text{gas}$ relationship. However, such models fail to reproduce the data completely, suggesting that thermal pressure plays an important role in the disks of dwarf galaxies.
\end{abstract}


\section{Introduction}
Star formation, the process by which gas is converted into stars, plays a central role in the evolution of a galaxy. Empirical constraints on the relationships between the present day quantities of gas and star formation in galaxies may therefore form a crucial component in sub-grid recipes of hydrodynamic simulations and semi-analytical models. \cite[e.g.][]{Furlong2015,Popping2014,Kuhlen2012,Lagos2011,Fu2010}; which trace the formation of baryonic structures (galaxies) in dark matter halos. Attempts to study the relationship between the gas and star formation rate (SFR) have been long standing \citep{Schmidt1959}, however only in modern times could any such relationship be demonstrated \citep{Kennicutt1998}.  In his pioneering study, \cite{Kennicutt1998} found a strong correlation between the SFR surface density and the total gas surface density, fitting a power law relation to a sample of spiral and starburst galaxies of the form,

\begin{align*}
\Sigma_{\text{SFR}} = A\Sigma^N_{\text{Gas}}.
\end{align*}
Power law parametrisations of these quantities are referred to henceforth in this work simply as a star formation relation, or interchangeably, the Kennicutt-Schmidt relation (KS).

More recent studies achieved sufficient sensitivity and angular resolution to study the inner regions of Local Volume (d $\lesssim$ 10 Mpc) spiral galaxies on kpc scales. These resolved studies instead demonstrated strict correlation between the observed surface densities of molecular hydrogen \textit{H$_2$} and the SFR surface density whilst simultaneously demonstrating an anti-correlation with atomic hydrogen (\HI) above its saturation limit \citep{Kennicutt2007,Bigiel2008,Leroy2008}. These studies indicated that the star formation relation was universal only when considering the molecular gas and not the total gas, for which the derived slope varied within and between galaxies. Indeed \cite{Krumholz2009} successfully derived an universal local star formation relation, with significantly reduced scatter when considering a `free-fall' time $t_{\text{ff}}$
\begin{align*}
&\Sigma_{\text{SFR}} = f_{H_2}\epsilon_{\text{ff}}\frac{\Sigma_{\text{gas}}}{t_{\text{ff}}}
\end{align*}
where $f_{H_2}$ is the fraction of the molecular gas and $\epsilon_{\text{ff}}$ is a dimensionless star formation efficiency (SFE) scale factor (see \cite{Krumholz2014} for a recent review on the subject). Further studies have challenged the universality of even the molecular star formation relation law \citep[e.g. see discussions by][]{Shetty2014,Leroy2013}. Comprehensive theoretical models should therefore accurately describe the observed distribution of apparent gas depletion times as a function of various environmental conditions or other factors and not just simply the surface densities of gas and SF. Comparing the agreement of the available data to theoretical models with varying underlying assumptions will therefore provide greater insight than just examining the linear regressions.

While in the inner disks of Local Volume spirals, the \HI content was not found to correlate with SFR surface density, the reverse is true in the outer \HI dominated disks and low-mass late type galaxies \citep{Bigiel2010,Bolatto2011}. Pushing H$_2$ column densities from 10 \msun pc$^{-2}$ to 1 \msun pc$^{-2}$, \citet{Schruba2011} and \citet{Bolatto2011} showed the molecular surface density still correlated with the SFR surface density. These facts together imply that the molecular fraction in the Interstellar Medium (ISM) reaches a floor value of a few percent \citep{Krumholz2013}. \cite{Ostriker2010} and \cite{Krumholz2013} developed theoretical models in order to describe the levels of star formation observed in the the \HI-dominated regime, with varying degrees of success. The \cite{Krumholz2013} model successfully describes the outer disks of spirals and dwarf galaxies \citep{Bolatto2011,Bigiel2010,Koribalski2009}. However for disk averaged quantities derived from dwarf galaxies \citep{Wyder2009,Roychowdhury2014} and the resolved \cite{Roychowdhury2015} data set, the \cite{Ostriker2010} model provides better agreement. A key difference between the two models is that \cite{Krumholz2013} includes the effect of gas metallicity on the transition from atomic to molecular hydrogen, which is inspired by the premise that H$_2$ forms on the surface of dust grains. The \cite{Ostriker2010} model does not consider the chemical state of the gas, but only its thermo-dynamic state by imposing pressure balance between the different ISM phases. With different underlying theoretical assumptions, distinguishing between these two models will provide important insight into the processes which govern star formation in \HI dominated environments.

While resolved studies of dwarf galaxies have provided a wealth of precision data, taken together, these studies \citep{Roychowdhury2015,Bigiel2010} were constrained to only 27 dwarf galaxies of the 107 observed by \cite{Hunter2012} and \cite{Begum2008} compared with the $\sim$ 400 irregular galaxies in the Local Volume. Future \HI surveys, whilst very sensitive, are unlikely to resolve anything but the nearest dwarf galaxies \citep{Koribalski2012,Koribalski2008}. Despite this, multi-wavelength coverage of dwarf galaxies in the Local Volume is particularly good, \citep[see][and references therein]{Karachentsev2013}, often only the \HI resolution is lacking. In order to increase the sample size to include dwarfs at greater distances, we must find approximate methods for computing the surface gas density. In this work, we provide a procedure for approximating the surface density of the gas from global \HI parameters and precision stellar photometry. We test the accuracy of our method by using the \cite{Wyder2009} and \cite{Kennicutt1998} galaxies as control samples.

This paper is organised as follows: In Section \ref{Method} we describe the galaxy samples, catalogues and our computed parameters. In Section \ref{results} we present our results, analysis and discussion. Particularly, we describe the accuracy of our gas surface density estimation. We then compare our data to resolved observations. {Although a vertically offset population to the spirals, we demonstrate that the sample of low surface brightness galaxies, predominantly dwarfs, forms a smooth distribution. Considering only the vertical offset and assuming a constant time-scale to convert the H$_2$ into star formation, we provide an estimation of a typical molecular fraction.} In Section \ref{modelsec} we compare all the available data in the literature to various models. Finally, we conclude in Section \ref{summary}.

\section{Methodology} \label{Method}
\subsection{Samples}

To study the star formation relation, (from here on we use the phrases star formation relation and the KS relation interchangeably) in a comprehensive sample of Local Volume dwarfs (M$_*$ $<$ 9 $M_{\sun}$), low-mass late types and low-surface brightness galaxies, we employ data from various sources in the literature which we describe below.

\subsubsection{Near-infrared photometry}
The global disk-averaged quantities used in the KS relation are typically computed from some fiducial radius corrected for inclination that is ideally representative of the star-forming disk. The choice of radius has not remained consistent over time. The K98 sample parameters were computed using the `RC2 radius', the \textit{B}-band 25th surface brightness isophotal diameters from the second reference catalogue \citep[i.e. $D_{25}$,][]{DeVRC2}. \cite{Wyder2009} instead used the circle defined by the semi-major axis of the ellipse used to extract the UV flux. \cite{Roychowdhury2014} has assumed that the star-forming disk is well described by the ellipse defined by the Holmberg diameter \citep{Holmberg1958}, which corresponds to a \textit{B}-band isophotal brightness of 26.5 mag arcsec$^{-2}$, based on morphological comparisons of the FUV, \HI and H$\alpha$. Photometric studies of nearby galaxies \citep[e.g.][]{Lauberts1989,DeVaucouleurs1991,Bremnes1998,Bremnes1999,Bremnes2000,Jerjen2000a,Barazza2001,Parodi2002,Makarova2005} were typically conducted in the optical regime and thus definitions of the stellar and star-forming disks naturally followed from \textit{B}-band photometry. Improvements in CCD technology and the introduction of near-infrared (NIR) detectors have since resulted in accurate \textit{JHKs} photometery of Local Volume galaxies \citep[e.g.,][]{Noeske2003,Vaduvescu2005,Vaduvescu2006,Vaduvescu2008,Kirby2008,Fingerhut2010,DeSwardt2010,McCall2012,Young2014}. In cases where the integration times are sufficient to overcome the bright sky background, the NIR offers several advantages over the optical when determining geometric properties of the stellar disk. It is significantly less sensitive to dust which will invariably attenuate and distort the flux \citep{Driver2007} and in addition the NIR flux contribution of the young stellar population is only significant in moderately strong starbursts \citep{Krueger1995}. Irregularities in morphology, such as randomly distributed \HII regions and associated H$\alpha$ emission are therefore less significant, allowing a greater accuracy in determining the stellar galactic center (and the resulting ellipse) which might have otherwise been centered on an offset H$_2$ region. 

When computing the parameters required for the KS relation, determinations of the disk radius employ the surface photometry from recent NIR Studies \citep[][Hereafter YJLK14, MVPB12, KRDJ08 respectively]{Young2014,McCall2012,Kirby2008} which have precisely measured the stellar disks of various Local Volume galaxies, mostly consisting of dwarf irregulars, low-mass late types, and blue compact dwarf galaxies. Using NIR surface photometry is unusual in the sense that we could simply use the Holmberg diameters from \citet{DeVaucouleurs1991} and \citet{Lauberts1989}. It is expected that the 'equivalent' Holmberg radii in the NIR will still be more accurate, despite requiring the \textit{B}-band magnitude to compute this parameter (see Section \ref{geomcomp}), for the reasons listed above.

In this study, the samples of YJLK14 and KRDJ08 are amalgamated to form the primary sample of galaxies since we have full access to their data and computed parameters. Together they form a sample of 79 galaxies with several properties. Firstly, a morphology which ranges from some low-mass ellipticals and high-mass lenticulars to many low-mass late types and dwarf irregulars. A broad range in stellar mass, 6.5 $<$ $\log_{10}(\mathcal{M}_{*}/\mathcal{M}_{\odot})$ $<$ 11 , with a sample median of 8.3. This sample is mostly dominated by low-mass late type galaxies and dwarf irregulars.  Finally, the KJRD08 and YJLK14 samples conveniently trace the main
cosmic structures of the Southern hemisphere out to 10 Mpc (the Sculptor and Cen A group, respectively).

To further supplement the YJLK14 and KRDJ08 samples, we also source data from the MVPB12 study. This is an amalgamated \textit{K}$_s$-band data set of newly observed galaxies and previous photometric studies filtered to include only those galaxies for which a surface brightness profile was successfully fitted, and for which the tip of the red giant branch (TRGB) distance was reliably measured. In addition to their own observations, MVPB12 sourced galaxy photometry from \citet[][34 galaxies]{Vaduvescu2005}, \citet[][17 galaxies]{Vaduvescu2008} and \citet[][80 galaxies]{Fingerhut2010}. The MVPB12 sample contains a total of 66 star-forming dwarf irregulars, which we analyze in addition to the dwarf irregulars and low-mass late types found in the YJLK14, KRDJ08 samples for a total sample size of 145 Local Volume galaxies.

\subsubsection{\HI fluxes, gas densities}
In ideal circumstances, suitably resolved \HI data of dwarf irregulars would be used to measure the gas densities point by point, and compared to the associated SFR surface density such as has been conducted in the studies of \cite{Roychowdhury2015} and \cite{Bigiel2008}. For dwarf galaxies, sensitivity and angular resolution of the \HI maps can be a particularly constraining factor (although the availability of SFR tracers may be an equally constraining factor), and so other previous studies have measured surface densities averaged over the entire star-forming disk \citep{Wyder2009,Roychowdhury2014}. Most galaxies in the YJLK14, KRDJ08 and MVPB12 samples described above, do not have readily available resolved \HI maps. In order to increase the statistics of dwarf galaxies in the make-up of the star formation relation diagram, we resort to deriving the atomic gas density from the available total \HI fluxes, averaged within the geometric parameters derived from the NIR photometric samples. 

We obtain \HI fluxes, with a few exceptions, from the \HI Parkes All Sky Survey (HIPASS) catalogues of \cite{Koribalski2004} (1000 Brightest Galaxy Catalog) and \cite{Meyer2004}. The HIPASS sample was observed on the 64m Parkes radio telescope using the 21cm multi-beam reciever, a correlator bandwidth of 64 MHz divided into 1024 channels and beam-width of $\sim$15 arcmin \citep{StaveleySmith1996}. The MVPB12 study has already compiled \HI fluxes for their galaxies from other sources (see their Table 4) and we simply adopt their values for those galaxies. For other galaxies we have obtained \HI data from either the \citet{Bouchard2005} or \citet[][Faint Irregular Galaxies GMRT Survey - FIGGS]{Begum2008} study. 

It is possible that the derived gas surface densities would introduce a bias and slope offset in the data which will require correcting. Potential statistical biases resulting from this estimation is explored in Sect. \ref{estimation}.


\subsubsection{UV and 24$\mu$m fluxes, SFR tracers}

Studies comparing the SFR estimates derived from \Ha$ $ and UV fluxes for Local Volume galaxies have found fundamental discrepancies where in principle it is expected they should agree \citep{Meurer2009,Lee2009a,Karachentsev2013a}. H$\alpha$ flux is systematically lower relative to the Far-Ultraviolet (FUV) with decreasing luminosity, underestimating the flux relative to the FUV by up to an order of magnitude. One explanation as \cite{Meurer2009} suggests, is that variations in the initial mass function (IMF) of low luminosity galaxies are responsible. This follows logically from the relative sensitivities of each tracer with respect to the IMF. H$\alpha$ emission resulting from ionization of the surrounding ISM will require stars in excess of 15 M$_{\odot}$ whose ionizing flux is sufficiently strong. In contrast, ultraviolet emissions directly trace the photospheric emissions of stars of several solar masses \citep{Kennicutt2012}. Other works by \citet{Fumagalli2011} and \citet{Weisz2012} instead suggest that temporal variation in the SFRs as the cause. 
{The above mentioned stochastic effects pronounce the gap between the underlying and traced star formation rate,}
appear to affect H$\alpha$ more significantly than FUV \citep{daSilva2014,Roychowdhury2014}. We therefore elect to use FUV fluxes when computing the SFR surface density. 

\begin{figure}
\center
\includegraphics[trim = 0mm 12mm 0mm 0mm, scale=0.55, clip=True]{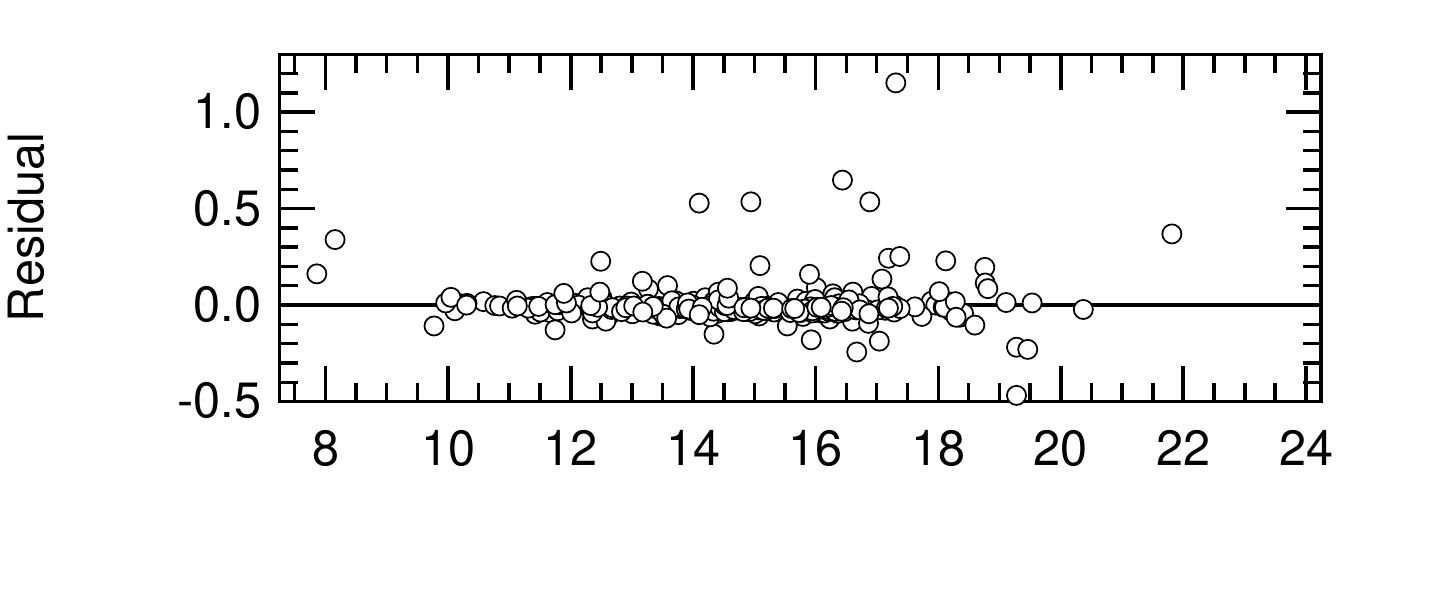}
\includegraphics[trim = -4.5mm 0mm 0mm 0mm, scale=0.55, clip=True]{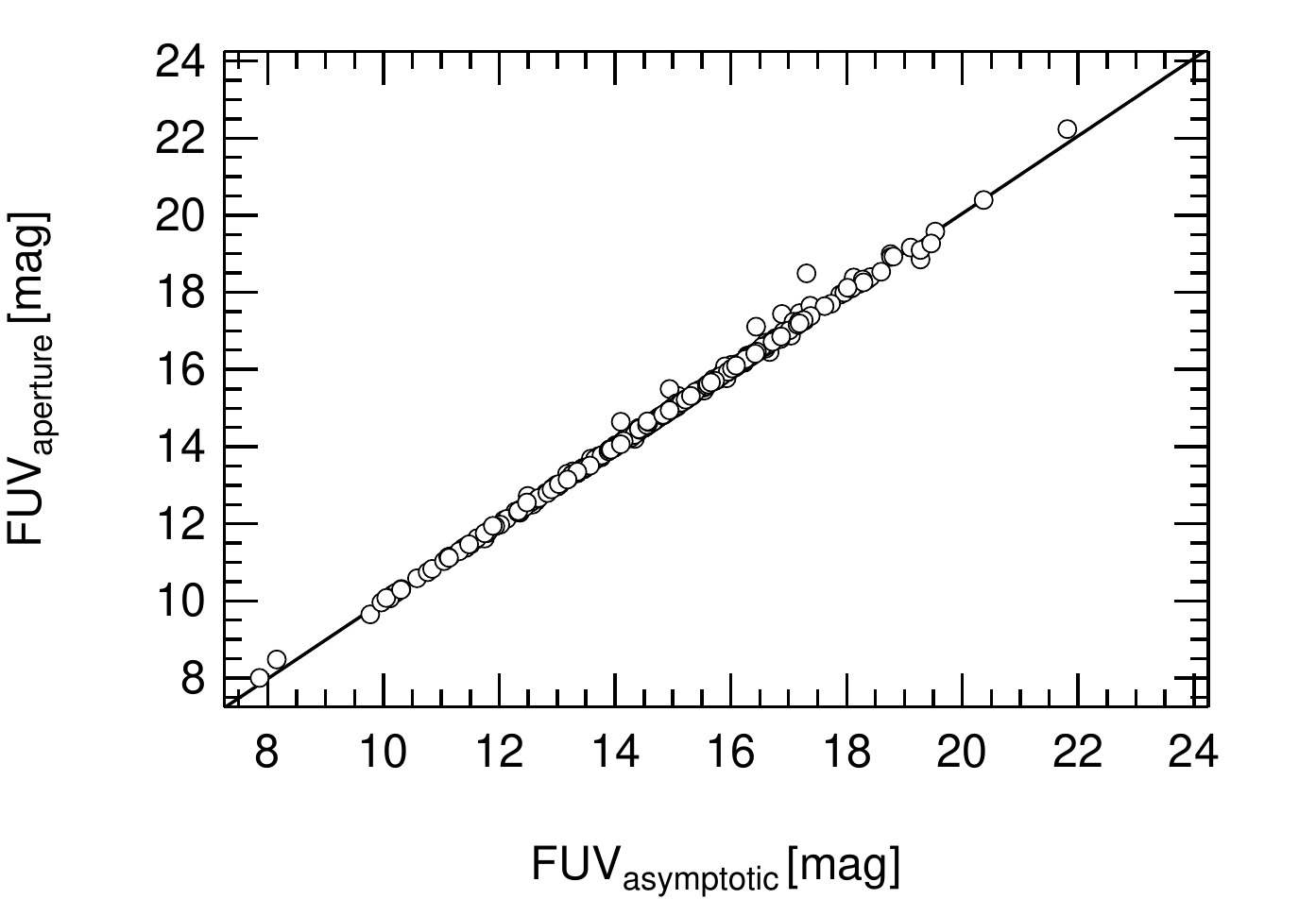}
\caption{Comparing the aperture and asymptotically derived total FUV magnitudes from the 11HUGS survey.}
\label{apvas}
\end{figure}

As part of the 11Mpc H$\alpha$ and Ultraviolet Galaxy (11HUGS) survey, \cite{Lee2011} presented UV photometry of a complete sample of Local Volume galaxies. We cross-correlate galaxies within the primary sample and MVPB12 extracting where available observed total FUV fluxes. In instances when a total FUV flux is not available, we instead use the aperture flux, defined as the aperture beyond which the flux error becomes 0.8 mag or where the intensity falls below the sky background level. Statistically, systematic differences between the aperture and total fluxes are negligible as shown in Fig.~\ref{apvas}, especially when compared to other sources of systematic errors.

The major drawback of using the FUV flux is its sensitivity to dust absorption. A composite tracer \citep[e.g.][]{Hao2011} accounting for the stellar emission re-radiated in the infrared via dust will therefore provide a better estimate of the SFR, although it is expected that for our sample dominated by low metallicity dwarf galaxies, these corrections should be small. The Local Volume Legacy survey \citep{Dale2009} is a legacy Spitzer Space Telescope volume limited survey designed to complement the 11HUGS and ACS Nearby Galaxy Survey Treasury \citep{Dalcanton2009} with observations of the near, mid and infrared fluxes of nearby galaxies. Using the 24 micron fluxes, we are able to correct our SFR tracer dust extinction in $\sim$35\% of our galaxies (for median change in the measured SFR of 0.04 dex). 

\subsection{Derived Parameters}
In the following subsections we describe the methods used to obtain the SFR surface density and the atomic gas surface density. Table~\ref{sampleq_appendix} compiles the sample observed properties while Table~\ref{sampled_appendix} lists the derived properties.

\subsubsection{Estimating the SFR}
\begin{figure}
\center
\includegraphics[trim = 0mm 0mm 0mm 0mm, scale=0.5, clip=True]{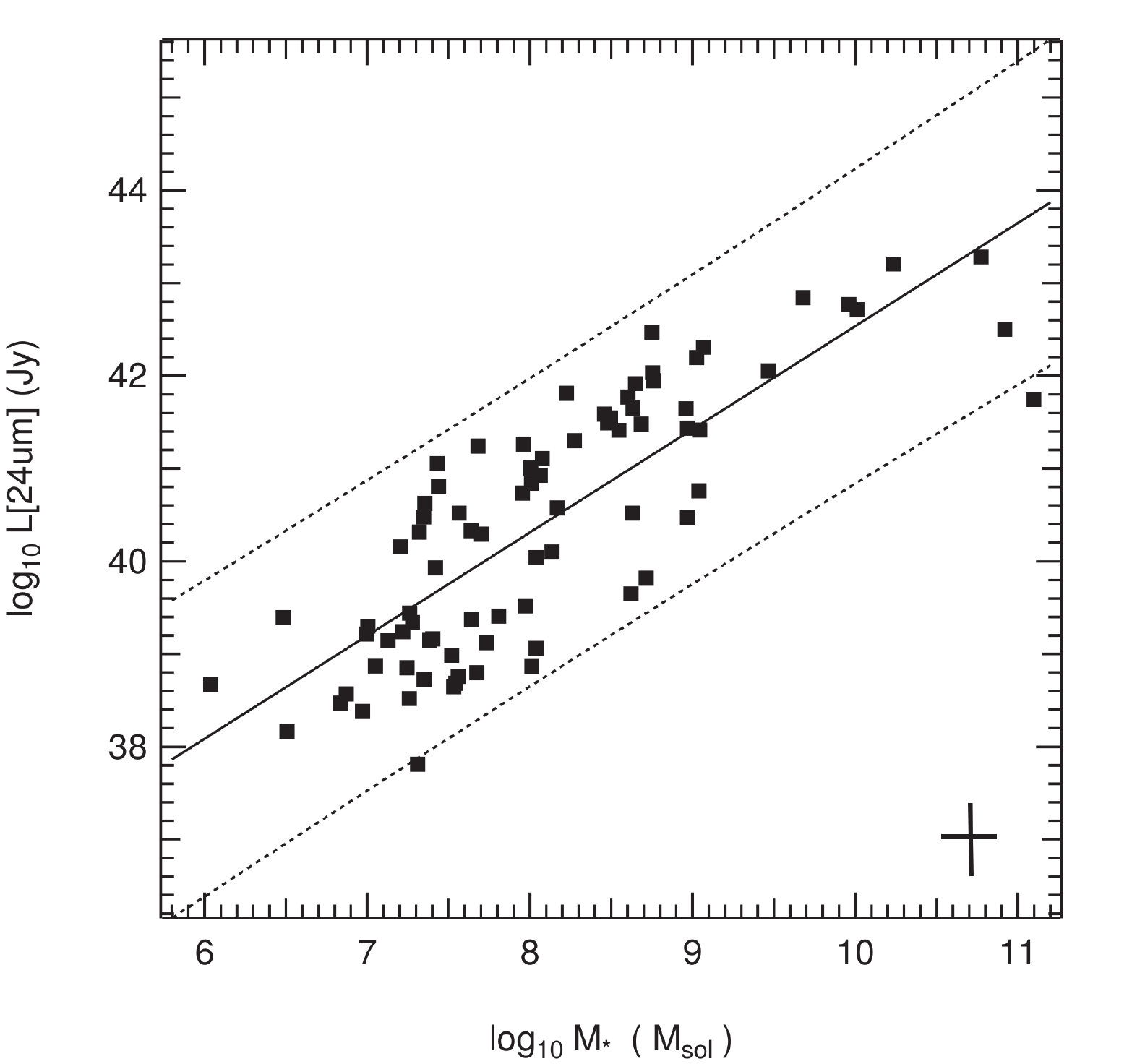}
\caption{The relationship between the stellar mass and 24$\mu$m luminosity in YJLK14, MVPB12 and KRDJ08 sample galaxies. The solid line represents the least squares fit to the data. The dotted lines represent the upper and lower 2$\sigma$ intervals. These intervals provide reasonable lower and upper bound estimates to the dust content as a function of stellar mass. The cross at the bottom-right shows representative $1\sigma$ errorbars.}
\label{dust}
\end{figure}

For estimating the \textit{FUV} traced SFR, we first correct the available asymptotic magnitude for Galactic extinction \citep{Schlafly2011} using the relation given by \citet{Karachentsev2013},
\begin{align}
m_{0,FUV} = m_{FUV} - 1.93 \, (A^G_B + A^i_B)
\end{align}
where $A^G_B$ and $A^i_B$ are the Galactic and internal extinction corrections respectively in the \textit{B}-band. We make no attempts to correct for internal extinction at this stage, since we account for internal extinction later using the 24 micron flux. The FUV flux density ($f_\nu$) is computed from the total FUV magnitude ($m_{FUV,0}$) given in Column~7 of Table \ref{sampleq_appendix},
\begin{align}
f_\nu (\text{erg} \text{ s}^{-1} \text{ cm}^{-2} \text{ Hz}^{-1}) = 10^{-0.4{\left(m_{FUV,0} +48.6\right)}}.
\end{align}
Using the best available distances in the literature (Column~5 in Table~\ref{sampleq_appendix}) and the flux densities, we compute the FUV and 24 micron luminosities ($L$), 
\begin{align}
L = \nu f_{\nu} \times 4 \pi D^2.
\end{align} 
%
For a constant star formation rate over 100 Myr, solar metallicity and either a Salpeter \citep{Salpeter1955} or a Kroupa IMF \citep{Kroupa2001}, the attenuation corrected FUV luminosity, $L(FUV)_\text{corr}$ (Column 2 in Table~\ref{sampled_appendix}), is given by \citet{Hao2011}:
\begin{align}
L(FUV,\text{corr}) = L(FUV) + 3.89 L(25\mu\text{m}).
\end{align}
The \cite{Hao2011} recipe was derived using the Infrared Astronomical Satellite (\textit{IRAS}) 25$\mu$m observations. We instead use the \textit{Spitzer} 24$\mu$m flux  (Column~8 in Table~\ref{sampleq_appendix}), which introduces an uncertainty of less than 0.3 mag \citep{Hao2011} due to the systematically lower luminosities recovered by IRAS observations \citep[][Fig.~1]{Kennicutt2009} relative to \textit{Spitzer}. 
Galaxies in the YJLK14, MVPB12 and KRDJ08 samples without a 24$\mu$m flux have their values estimated using the following relationship as a zeroth order approximation.
\begin{align}
\log_{10} L\left(24\mu\text{m}\right) = 1.11^{+0.09}_{-0.09} \log_{10} \frac{ \dot{M}_*} { \left[\text{M}_{\odot}\right] } + 31.415^{+0.739}_{-0.739}
\end{align} 
derived from the least squares regression of the galaxies with available 24$\mu$m flux measurements as shown in Fig.~\ref{dust}. As demonstrated, the 2$\sigma$ trends provided reasonable upper and lower bound limits to the available data. We have conservatively elected to use the lower bound relationship to correct for dust since it is plausible that many dwarf galaxies in the sample have inconsequential dust content. We note that dust measurements remain a large source of uncertainty for those galaxies without a 24$\mu$m flux measurement, since the data are scattered over approximately 2 orders of magnitude.

Plugging the corrected FUV luminosity into the recipes prescribed by \citet{Kennicutt2012} and \citet{Hao2011} we obtain the SFR ($\dot{M}_*$, Column~3, Table \ref{sampled_appendix}),
\begin{align}
\log_{10} \frac{ \dot{M}_*} { \left[\text{M}_{\odot} \text{ yr}^{-1}\right] }= \log_{10} \frac{L(FUV,\text{corr})}{\left[\text{erg s}^{-1}\right]} - 43.35.
\end{align}
As to be expected for the dwarf and LSB galaxies, dust corrections to the SFRs are typically low. The median change in the measured SFR is 0.04 dex although a maximum of 0.5~dex is recorded indicating the importance of dust corrections in the case of some galaxies.

Star formation rate values from \citet{Wyder2009} and \citet{Kennicutt1998} require a correction ($\dot{M}_* = 0.63 \dot{M}_{*,K98}$) to account for updated stellar population models and the choice of IMF used in more modern studies. We have performed this correction in the analysis to bring them into line with more modern studies.

We calculate the uncertainty in the measurements by including the uncertainty in distance (typically between $\approx 10-20$\%), the flux
and the errors in the calibration.
To get total errors, we add these uncertainties in quadrature. 
In all the figures we show representative error bars, which were calculated as the average error on the quantities of the galaxies shown.

\subsubsection{HI and gas mass estimation}

We estimate the \HI mass (compiled in Column~4, Table \ref{sampled_appendix}) for our sample galaxies following \citet{Roberts1975},
\begin{align}
M_{\HIsub} = 2.36 \times 10^5 D^2 F_{\HIsub},
\end{align} 
where D is the distance of the galaxy from the sun and $F_{\HIsub}$ is the 21 cm emission line flux (Column~9, Table \ref{sampleq_appendix}). 
The dwarf and low-mass late type galaxies are assumed to have a negligible molecular fraction and so the total
gas mass $M_\text{gas}$ is simply the \HI mass corrected for the presence of helium using a factor of $1.34$.

\subsubsection{Geometric and density computations} \label{geomcomp}

In order to find the NIR equivalent Holmberg radius $R_{26.5,eq}$, we use the extinction corrected \textit{(B-H)$_0$}, and \textit{(B-Ks)$_0$} colours.  YJLK14 demonstrated (their Figure 10) significant variation in the colours of dwarf galaxies and thus it is important to perform colour corrections directly rather than through the use of a scaling relationship. 


The NIR equivalent Holmberg radius ($R_{26.5,eq}$) for the primary sample is computed from Equation 1 in  YJLK14 and is given by
\begin{align}
R_{26.5,eq} &= \left[0.921(26.5 - [(B-H)_0])-0.921\mu_0\right]^{1/n}\,r_0,
\end{align}
where $\mu_0$ is the central surface brightness in the \textit{H}-band, $n$ is the S\'{e}rsic index and $r_0$ is the scale length parameter in units of pc. Similarly, following from Equation 1 in MVPB12, 
\begin{align}
R_{26.5,eq} &= \text{acosh}\left[10^{\left[-0.4\left(\mu_0 - (26.5 - [(B-Ks)_0])\right)\right]}\right]\,r_0.
\end{align} 

For the amalgamated MVPB12 sample, all parameters are generated using the semi-major axis as the native radial coordinate and so no further corrections are required to compute the area of the disk defined by the semi-major axis. The KRDJ08 and YJLK14 data was instead computed using the geometric mean radius. The area defined by the semi-major axial component of the $R_{26.5,eq}$ in this case is,
\begin{align}
A = A_{26.5,eq} = \pi R^2_{26.5,eq} / (1-e),
\end{align}
where $e$ is the measured ellipticity. The estimated SFRs and \HI masses are averaged over this quantity to derive an estimate of the atomic gas and SFR surface densities for the NIR galaxy samples. Along with $R_{26.5,eq}$, the quantities are compiled in Table \ref{sampled_appendix}. The \cite{Tully1988} catalogue include an inclination corrected Holmberg diameter and it is therefore trivial to calculate the atomic gas and SFR surface densities.

\section{Results and Discussion} \label{results}

\begin{figure}
\center
\includegraphics[trim = 0mm 0mm 0mm 0mm, scale=0.32, clip=True]{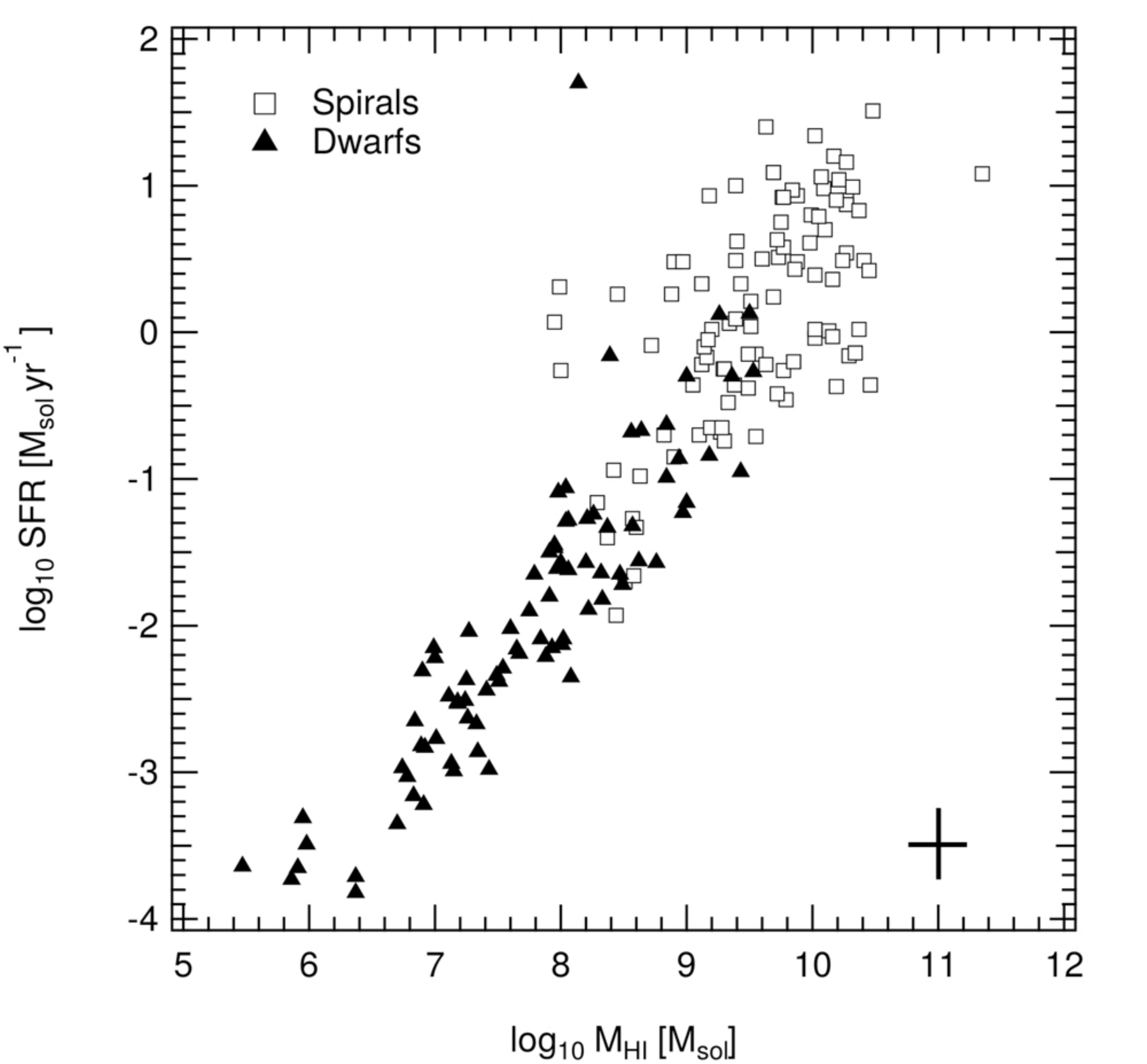}
\caption{The relationship between the SFR and total \HI mass in dwarf and spiral galaxies. Data for the dwarf galaxies were drawn from YJLK14 and KRDJ08. Data for the spiral galaxies were obtained from the NHICAT, HOPCAT and ALFALFA \HI catalogues. Dwarf galaxies show a very consistent and linear relationship over a broad range of \HI masses, but spiral galaxies display comparatively more scatter than dwarfs. The cross at the bottom-right shows representative $1\sigma$ errorbars.}
\label{SFRvHI}
\end{figure}

The main focus of the following discussions is to investigate the trends in the star formation relation in the \HI dominated regime. Noting the limitations a linear model pose on an obviously multidimensional problem, a later section is dedicated to comparing the gathered data to various comprehensive models of star formation.

\subsection{How accurate is $M_{\HIsub}/A$ as a proxy for $\Sigma_{\text{gas}}$?} \label{estimation}

Before proceeding, we must ensure that $M_{\HIsub}/A$ is a valid proxy for $\Sigma_{\text{gas}}$. The apparent correlation and linearity between the total atomic gas and the star formation rate as shown in Fig.~\ref{SFRvHI} suggest this might be the case. Dwarf galaxies show a very consistent and linear relationship over a broad range of \HI masses. The linearity of this relationship suggests that by normalizing to a fiducial disk area, such as that defined by the Holmberg diameter, we do not introduce a secondary effect on the measured slope of the linear regression to the data points. Of course, we may introduce a {bias and an offset} in the $\Sigma_{\text{gas}}$ scale since by using the $M_{\HIsub}/A$ proxy we are insensitive to variation in the ratio of the optical-\HI diameter ($A$ is derived following the optical data).

\begin{figure}
\center
\includegraphics[trim = 0mm 0mm 0mm 0mm, scale=0.5, clip=True]{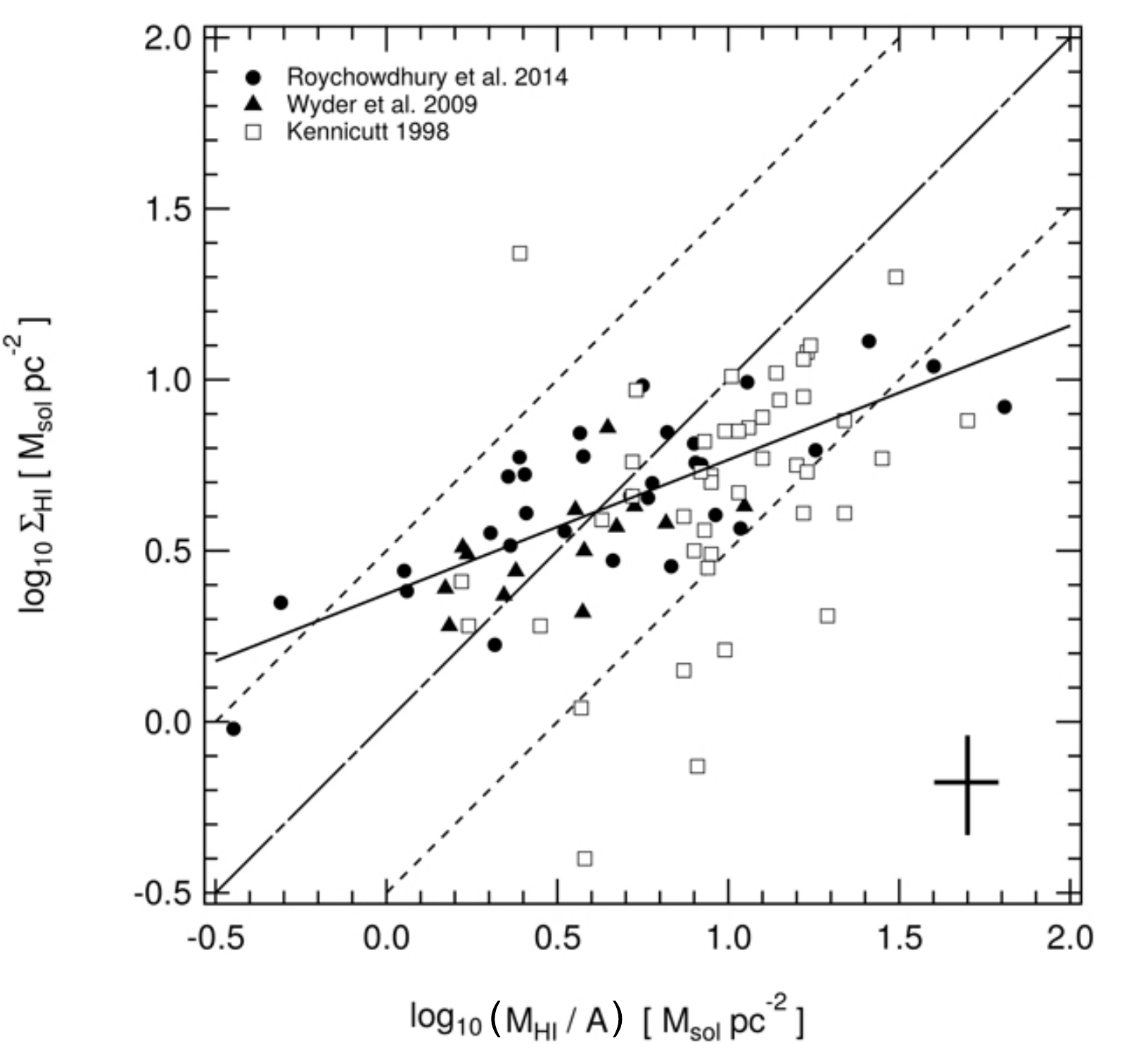}
\caption{$\Sigma_{\text{\HIsub}}$ versus $M_{\HIsub}/A$ for the K98 spiral and \protect\cite{Wyder2009} and \protect\cite{Roychowdhury2014} dwarf sample galaxies. {The broken solid line indicates a slope of unity. Dashed lines indicate offsets of 0.5 dex. The solid line shows the least squares regression to the low surface brightness galaxies from the \protect\cite{Wyder2009} and \protect\cite{Roychowdhury2014} samples (correlation coefficient 0.750).} The cross at the bottom-right shows representative $1\sigma$ errorbars.}
\label{proxyvSigma}
\end{figure}

In order to determine the typical value by which we over or under-estimate the actual gas surface density, we calculate $M_{\HIsub}/A$ for the spiral galaxies from K98, and the dwarf galaxies from \citet{Wyder2009} and \cite{Roychowdhury2014} comparing them to the $\Sigma_{\text{gas}}$ calculated in the respective studies. As mentioned previously, we obtain global \HI fluxes, inclinations and isophotal radii for the K98 galaxies from the Nearby Galaxy Atlas catalogue. For the low surface brightness galaxies, these quantities are obtained from the RC3 catalogue and for the \cite{Roychowdhury2014} these are obtained from \cite{Begum2008}. Figure \ref{proxyvSigma} plots the surface gas densities against the computed $M_{\HIsub}/A$ quantities. 
We emphasize $A$ has been derived here not using the \HI data but the optical data, i.e., considering the Holmberg radius and adding the factor $1/\cos i$ to include the geometric correction (see YJLS14).
As shown in the plot, most data points are scattered around unity typically by within 0.5~dex. There is an obvious trend from overestimation to underestimation with increasing \HI mass. Fortunately this bias appears to be linear and we correct for it using a least squares approach. The Pearson correlation coefficient of this fit is $r = 0.750$. We derive the correction  
\begin{align}
\log_{10} \Sigma_{\HIsub} =  (0.4\, \pm\, 0.1)\log_{10} ( {\rm M_{\HIsub}}/A_{26.5} ) + (0.38\, \pm\, 0.04), 
\end{align}
(solid line in Fig.~\ref{proxyvSigma}), which is applied before correcting for helium. 
We include the errors in the fit above when estimating the total errors in $\Sigma_{\rm HI}$ and $\Sigma_{\rm gas}$.

From Fig~\ref{proxyvSigma} and the demonstrated large scatter in the K98 sample, we recommend not to use this method in galaxies other than \HI-dominated galaxies. 
Given that  many dwarf galaxies, particularly those with distances larger than 11~Mpc, will remain unresolved for the foreseeable future  --large \HI surveys using the ASKAP and MeerKAT will typically have a beam size of $\sim$30", which is just 2--6 times smaller than the typical size of the \HI distribution in dwarf galaxies of the Local Volume, \citep[e.g.][]{Koribalski2018}, and hence only the closest dwarfs will be resolved in \HI--,
this method is the only way to compare dwarfs to spirals in statistically meaningful samples and it is therefore fortunate that the scatter for dwarf galaxies in Fig.~\ref{proxyvSigma} is within an order of magnitude and the bias is linear.

\subsection{The Kennicutt-Schmidt relation for Local Volume galaxies}

\begin{figure}
\center
\includegraphics[trim = 1mm 0mm 0mm 0mm, width=0.47\textwidth, clip=True]{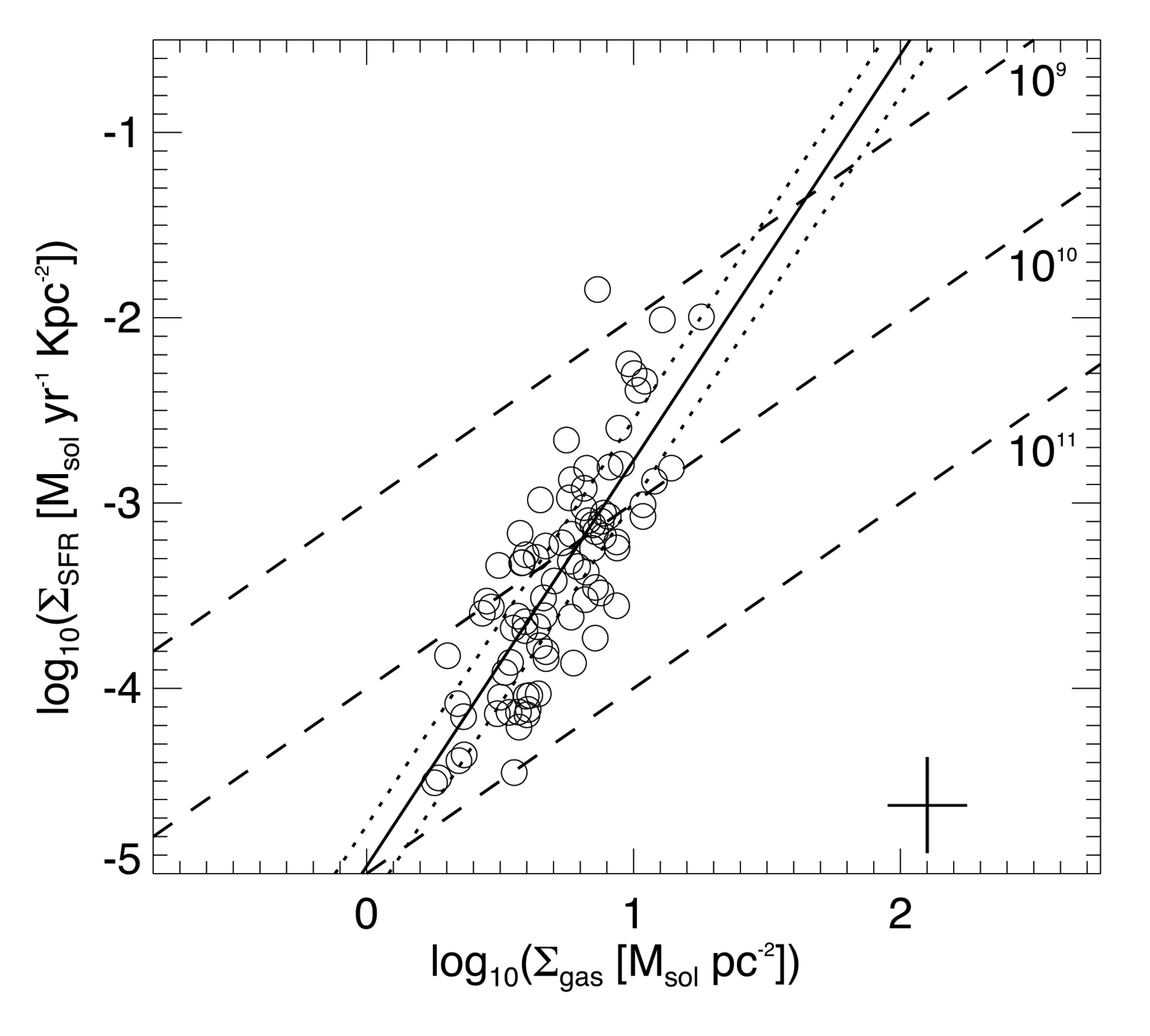}
\caption{{Surface density of the SFR as a function of the gas surface density for the low surface brightness galaxies obtained from the NIR samples, KRDJ08, YJLK14 and MVPB12 and whose atomic gas surface densities have been estimated from global parameters as outlined in Sect~\ref{Method}. Dashed lines corresponds to slopes of constant gas depletion times in units of years, as labelled. The solid black line corresponds to the best-fit 
obtained using {\tt HYPER-FIT} \citep{Robotham15}, adopting the conjugate gradients method. The dotted lines mark the $1\sigma$ uncertainty in the 
zero-point of the relation. The cross in the bottom-right of the panel shows a representative error bar of individual measurements.}}
\label{KSlawdwarfs}
\end{figure}

\begin{figure}
\center
\includegraphics[trim = 1mm 0mm 0mm 0mm, width=0.47\textwidth, clip=True]{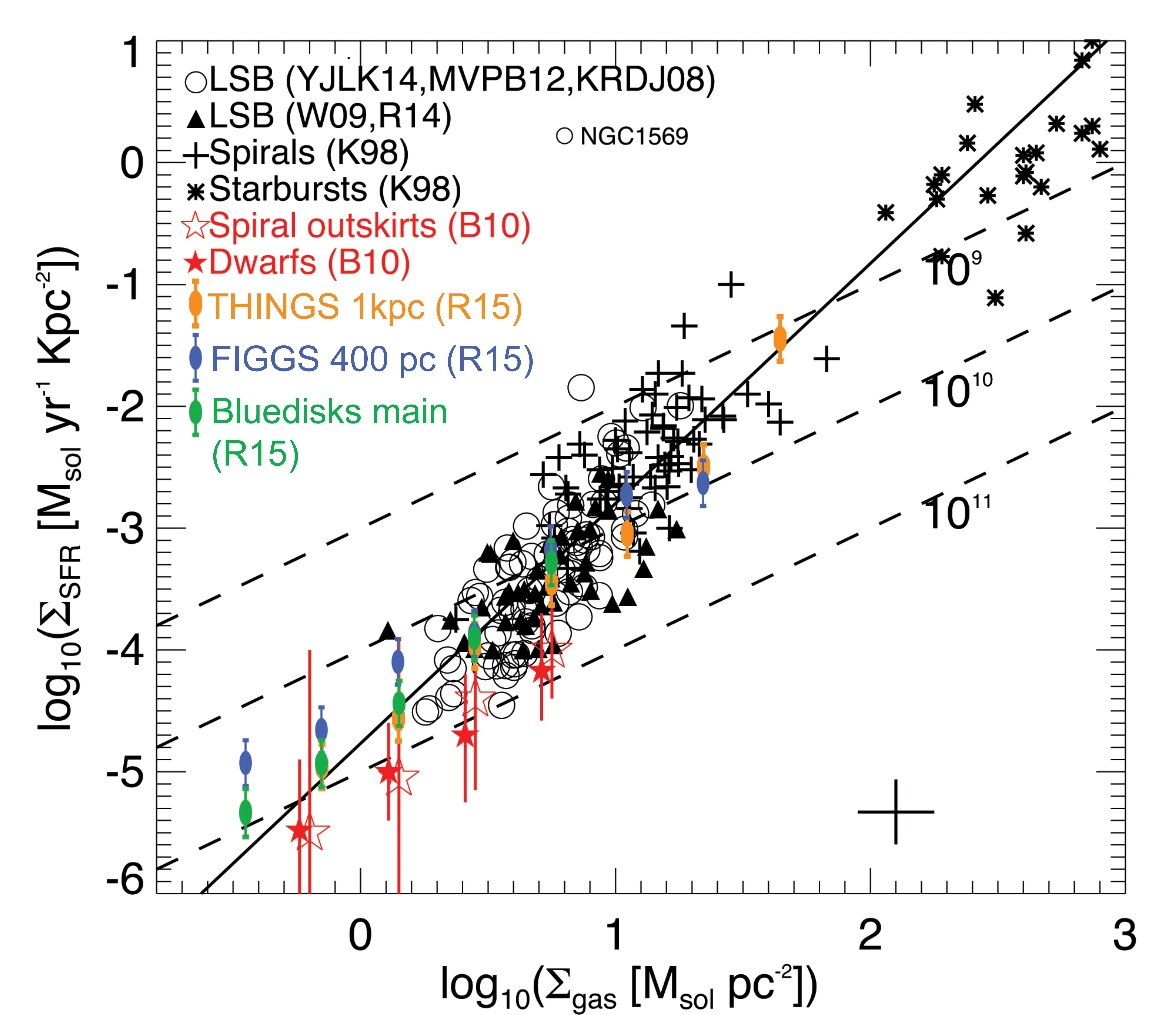}
\caption{The Kennicutt-Schmidt relation for YJLK14, MVPB12 and KRDJ08 (\textit{open circles}) sample galaxies compared to the original K98 data (\textit{plus symbols} for spirals, \textit{crosses} for starbursts). Additionally, we have included the disk averaged quantities for the galaxies whose column densities were measured directly \protect\citep[][\textit{black triangles}]{Roychowdhury2014,Wyder2009}. 
The open and filled stars with errorbars show the median and $1\sigma$ scatter of the spiral outskirts and dwarf galaxies of \protect\cite{Bigiel2010}, respectively, both of which correspond to \HI-dominated regions. For clarity, we artificially displaced the medians of the dwarf galaxies by $0.04$~dex in the $x$-axis. 
The \protect\cite{Roychowdhury2015} data for their {'THINGS~1kpc' (orange circles),} 'FIGGS~400pc' (blue circles) and 'Bluedisks main' (green circles) samples are also shown for comparison.
The K98 galaxies have been corrected for the presence of helium contrary to the original work. The diagonal dashed lines corresponds to slopes of constant gas depletion times in units of years, as labelled. The solid line shows the best fit to selected data, as described in the text, using the same the procedure described in Fig.~\ref{KSlawdwarfs}. The cross in the bottom-right of the panel shows a representative error bar of individual measurements.}
\label{KSlaw}
\end{figure}

Using the data and derived parameters described in Section~\ref{Method}, we plot the gas and SFR surface densities. Figure~\ref{KSlawdwarfs} explores the simple linear regression to the SFR relation for low surface brightness galaxies from KRDJ08, YJLK14 and MVPB12. Figure~\ref{KSlaw} extends the regression analysis to include low surface brightness galaxies and spiral galaxies from further sources \citep{Roychowdhury2014, Wyder2009, Kennicutt1998} whilst the outer disk \citep{Bigiel2010,Roychowdhury2015} and starburst data \citep{Kennicutt1998} is included in the plot for comparative purposes only. 

In order to obtain a best-fit to our data and quantify the power-law index of the relation $\Sigma_{\rm SFR}\propto \Sigma^{N}_{\rm gas}$, 
we use the {\tt HYPER-FIT} R-package of \citet{Robotham15}. We perform the fits using the Optim algorithm with the conjugate gradients method, 
minimising the scatter orthogonal to the plane. In the fit we included the errorbars for individual objects, and assume that that the errors 
in the $x$ and $y$ axis are not correlated. The best fit obtained for the low surface brightness galaxies from KRDJ08, YJLK14 and MVPB12 is
\begin{align}
\log_{10} \Sigma_{\text{SFR,LSB}} = 2.19^{+0.25}_{-0.21} \log_{10}\Sigma_{\rm gas} - 4.96^{+0.22}_{-0.22}
\end{align}
including all the data as outlined above, the corresponding best fitting relation is instead,
\begin{align}
\log_{10} \Sigma_{\text{SFR,spirals and LSB}} = 1.97^{+0.11}_{-0.1} \log_{10}\Sigma_{\text{\HIsub}+H_2} - 4.77^{+0.28}_{-0.28}
\end{align}

Figure~\ref{KSlawdwarfs} shows that there is a clear correlation between the surface densities of the SFR and the atomic gas with a steep dependence on the latter quantity. Although we did not include the \cite{Bigiel2010} data in the least squares fit shown in Fig.~\ref{KSlaw}, visual comparison of their data set to the line of best fit seems to show a good agreement, although we remind that the  \cite{Bigiel2010} points have too high depletion time.
We note, however, the narrow range in $\Sigma_{\text{\HIsub}}$ for our sample galaxies, which spans from 0.2 to 1.2 in log scale in Fig~\ref{proxyvSigma}. This is a consequence of the fit to the data from \citet{Wyder2009} and \cite{Roychowdhury2014} samples that we apply to the unresolved HI data.
Possible biases in the samples, for example by being sensitive to only a narrow range in $\Sigma_{\rm SFR}$ or $\Sigma_{\text{\HIsub}}$  \citep[unlike the spiral galaxies from][which span a much wider range in both axes]{Kennicutt1998}, could lead to the data preferring an N=1 slope due to the reduced dynamic range. If this were the case, using resolved \HI data to estimate \HI distributions could change our results.


The slopes derived for the LSB galaxies in our study are much steeper than that found in \cite{Roychowdhury2014} and is more supportive of the findings of \cite{Roychowdhury2015}. 
\cite{Roychowdhury2015} suggested this difference could be an effect of being biased towards the inner star-forming regions of galaxies. If that were the case, we might expect the same phenomenon in our sample galaxies which are also disk averaged quantities. Instead they form a tight distribution intermediate to the outer disk data of \cite{Bigiel2010} and the spirals of \cite{Kennicutt1998}. 
Indeed, we note that the two determinations of the K-S relation slope (1.97 for the combined spirals+LSB sample and 2.19 for the LSB sample) agree within the fitting errors. Therefore, using the LSB sample we are probably circumventing any existing bias (as the mentioned narrow range in $\Sigma_{\text{\HIsub}}$) in the \citet{Wyder2009} and \cite{Roychowdhury2014} samples and reinforcing the results obtained by \citet{Kennicutt1998}.

Our sample includes galaxies over a much wider range of surface brightness than the sample of \cite{Roychowdhury2014}: low surface brightness galaxies from YJLK14, the somewhat brighter low-mass late types and irregulars from KRDJ08, as well as star-forming dwarfs and additional low surface brightness galaxies from the amalgamated MVPB12 sample. We suggest that N=1 slope would be simply due to insufficient statistics especially given the generally good agreement between the \cite{Roychowdhury2014}, \cite{Wyder2009} the novel KRDJ08, YJLK14 and MVPB12 datasets. For dwarf galaxies, we find a slope that is strictly not in agreement with the canonical N=1.4 derived from the original K98 sample. As \cite{Bigiel2008} noted, the N=1.4 relies on the contrast of spiral disks and the molecular rich circumnuclear starbursts.


Notably, the galaxy NGC~1569 (labelled in Fig.~\ref{KSlaw}) lies well outside the expected behavior for spirals or dwarf galaxies. Environmental interactions resulting in the star-burst activity are very likely to be the cause of NGC 1569 significant offset from the star formation relation with respect to the broader population of galaxies \citep[e.g.][]{Muhle2005}, as also discussed in the case of NGC~5253 \citep{LS10,Lopez-Sanchez2012}.

\subsection{Comparing global and resolved properties} \label{gvc}

Above the critical density of the total gas ($\sim$ 9 \msun pc$^{-2}$), in the H$_2$-dominated regions of their spiral galaxies, \cite{Bigiel2008} and \cite{Leroy2013} found an universal molecular star formation relation with a corresponding exponent of N=1. Conversely, when looking at the corresponding total gas or \HI gas star formation density diagrams, exponents are found to vary significantly radially within galaxies and from galaxy to galaxy. Above the critical density, the atomic gas saturates in these galaxies and there is no observed correlation between the atomic or total gas with the SFR. While, below the critical density where the gas is \HI-dominated, the atomic gas and SFR surface densities correlate \citep{Bolatto2011,Bigiel2010,Koribalski2009,Lopez-Sanchez2015}. This relationship does not have a fixed slope. 

Variation of the total gas SFR relations imply that the ratio of \HI-to-H$_2$ can vary within galaxies and in between galaxies, or rather that the star formation efficiency (SFE) is not constant due to either local or global factors.  More recent work has confirmed significant variation of the SFE between different THINGS galaxies \citep{Shetty2014}, but they studied the molecular hydrogen.
While factors other than the surface density of the gas have been shown to influence the SFE, in their sample, \cite{Leroy2008} demonstrated where the ISM is dominated by atomic gas, such as in the outer disks of spirals, the SFE decreases with increasing radius, implicating a global radial dependence on SFE in addition to local variation. The studies of \cite{Roychowdhury2015} and \cite{Elmegreen2015} provide further evidence for radial dependance on the SFE. 

Recently \cite{Wang17} used high-quality interferometric \HI data from the LVHIS \citep[Local Volume \HI Survey][]{Koribalski2018}, 
including a multi-wavelength dataset that allows to perform a careful estimation of the SFR, to
find that the correlation between the globally averaged $\Sigma_{SFR}$ and $\Sigma_{\rm H\,I}$ is weak. 
Their figure~11 shows these data; their galaxies seem to overlap on the scatter from our Fig.~\ref{KSlaw}. 
They also found that the SFE significantly depends on the average stellar surface density, something that can be explained using a marginally stable disk model, as described in \cite{Wong16}.  This was already explored by \citet{Shi2011}, who proposed an explicit dependence of the SFE on the stellar mass surface density (the "extended Schmidt law"). The SFE-$\Sigma_{\rm star}$ relation proposed by \citet{Shi2011}, that can be reproduced by some models and holds over five orders of magnitude in -$\Sigma_{\rm star}$ for individual global galaxies, included LSB and dwarf galaxies that deviate from the K-S law.

With the increase in sample size relative to prior studies, we are able to reproduce similar trends shown in resolved studies and reconcile some of the inconsistency introduced by \cite{Roychowdhury2014}. Both resolved and disk averaged studies imply a `floor' on the fraction of H$_2$ gas given the shape of the distributions in Fig.~\ref{KSlaw}. The resolved and disk averaged data are not in complete agreement when it comes to the slope of the star formation relationship. \cite{Bigiel2010} and \cite{Roychowdhury2015} find exponents on average of N = 1.5 while this study finds a much steeper slope (N = 2.4) for the disk averaged data, albeit with a large uncertainty. This statistical uncertainty in the slope greatly decreases when including the spiral disk data from \cite{Kennicutt1998} and this is reflected in the correlation coefficients of these fits increasing from r = 0.670 to 0.795.

\cite{Roychowdhury2015} suggested that this apparent inconsistency arises from the bias when looking at optical disk averaged quantities. It is possible to imagine resolved regions within a \HI disk with similar gas surface densities to a disk averaged value for a given galaxy but with little or no star formation as they may lie outside the star forming disk. One potential means of resolving this inconsistency is to use the total area of the \HI disk rather than the optical disk. This would require a trend in the optical to \HI diameter such that galaxies with higher surface densities had a larger diameter ratio relative to galaxies with lower surface densities.

Ultimately as the distribution of points in Fig.~\ref{KSlaw} demonstrate, the star formation relationship is more complex than what linear relationships would imply. We note that our data occupy a narrow range of atomic surface densities, whilst the numerous resolved data points from \cite{Bigiel2010} occupy a much wider range and show a complex distribution. Within this distribution itself significant variation of the exponent of the star formation law is easily possible. We suggest that such variations in slope are an indicator of the complexity of the problem and that a more nuanced approach would be required to fully describe the distribution of the data. Section~\ref{modelsec} describes the available comprehensive models dealing with the star formation relationship and compares the available data to them in order to obtain greater insight.

\subsection{Physical causes for the variation in the SFE} \label{molfrac}

At least some of the scatter, and therefore the overall offsets in the distribution of the Local Volume galaxies, particularly the dwarfs, must occur from variations in the scale height of the disk. \cite{Schmidt1959} originally proposed that $\rho_{SFR}\sim\left(\rho_{\text{gas}}\right)^n$, whereby $\rho_{\text{SFR}}$ and $\rho_{\text{gas}}$ are the SFR and gas volume densities respectively. 
For constant scale-heights, such as in ideal spirals whose intrinsic axial ratio is well known ($\sim$ 0.2), projection effects play no part in the variation of SFE. There is evidence that Local Volume galaxies and dwarfs do not have a constant scale-height \citep{Roychowdhury2010,Roychowdhury2013}. 

Modeling a galaxy's stellar discs as triaxal ellipsoids, \cite{Roychowdhury2013} used Monte Carlo simulations to determine the minor ($q$) and major ($p$) intrinsic axial ratios of irregular galaxies of the Local Volume. 
These galaxies demonstrated a significant degree of variation in their intrinsic axial ratios \mbox{0 $<$ $q_0$ $<$ 1},
which was mildly correlated with luminosity. 
Perhaps of more relevance to this study is the intrinsic axial ratio of the gas distribution. For a sample of faint dwarf irregulars from the FIGGS sample, \cite{Roychowdhury2010} determined a mean axial ratio of $\langle q \rangle$ $\sim$ 0.6 for their sample. Note, however, that the axial ratio of $\langle q \rangle$ provided by  \cite{Roychowdhury2010} had a large associated error, and therefore a large intrinsic variation in the ratio from galaxy to galaxy is expected.

Assuming a disk in which the gas volume density decays exponentially, the surface density $\Sigma_\text{gas}$ and central volume density $\rho_{0,\text{gas}}$ is related by the scale height as \mbox{$\rho_{0,\text{gas}}$ $=$ $\Sigma_{\text{gas}}/(2h)$.} As in the LSB galaxies in the \cite{Wyder2009} and the \cite{Roychowdhury2014} sample, the Local Volume galaxies in this study also display a factor $\sim$5 offset in {gas depletion times} to the spiral in the original K98 sample, which if entirely attributable to changes in scale height, would imply that late type and irregular galaxies have discs with significantly larger scale heights relative to spiral galaxies. 

Relatively speaking, potentially thicker disk scale-heights in dwarf irregulars would lead to a lower effective volume density, increased free fall time and thus lower observed SFE. Indeed in their study of 20 dwarf irregular galaxies, \cite{Elmegreen2015} considered the effects of disk thickness extensively. The dwarf galaxies in their sample were shown to be thick in absolute terms, with gas scale heights of 0.5 kpc and a typical ratio of scale height to radius of 0.6. Remarkably, correcting for the free fall time of the atomic gas (see Eq. \ref{sflaw}) yields an efficiency $\epsilon_{\text{ff}}$ of one per cent, which is the also the SFE of the molecular gas. In other words the atomic gas is consumed at the same rate as molecular suggesting their ratios are fixed.

Dwarf galaxies are \HI-dominated and so the molecular fraction is expected to be low which also simultaneously leads to a lower effective SFE of the total gas. Measuring the molecular fraction in low metallicity galaxies is challenging, due to the very high ratio of CO-to-H$_2$ column densities \citep{Bolatto2013}, although CO observations \citep[e.g.][]{Schruba2012} and dust emission studies \citep{Bolatto2011} suggest that the molecular gas fraction should be very low. For example, \cite{Bolatto2011} derives a molecular fraction of only 5\% for the Small Magellanic Cloud (SMC) using high-resolution \HI and dust emission maps. Given that the SFE of the atomic gas and molecular gas are found to be similar, it should be possible to estimate the fraction of molecular gas by comparing the atomic and molecular star formation laws and having adjusted for free fall times (or in our case a disk thickness factor. 

Although evidence in general supports dwarf irregulars having thick disks \citep[e.g.][]{Roychowdhury2010,Elmegreen2015}, direct measurement of the three dimensional shape of the dwarf galaxies DDO~46 and DDO~168 using the central stellar dispersion to the ratio of the maximum rotation speed by \cite{Johnson2015}, suggest that these galaxies are comprised of thin stellar disk. 
If gas disks in dwarf galaxies were also thin, the estimation of the molecular fraction in these systems should also account for the possibility of varying disk thickness and free fall times.

To attempt an estimation of the typical molecular fraction we consider only variations in disk scale height and molecular fractions. We ignore other factors which might affect the SFE. First we assume that disk scale heights are largely consistent within the dwarf galaxy sample and following from \cite{Roychowdhury2010} the disk thickness is statistically $\sim$2.5 times thicker than the typical spiral. Using the relationship between molecular gas and SFR surface densities from \cite{Leroy2013}, our galaxies are offset by a factor of approximately 15-20 to the expected value. However since the disk is potentially 2.5 times effectively less dense than for those in spirals due to an increased scale height this factor is perhaps closer to 6-8 times corresponding to a typical molecular fraction of $\sim$ 0.16, or in the case of a thin disk 0.05.

\begin{figure}
\center
\includegraphics[trim = 1mm 0mm 0mm 0mm, scale=0.5, clip=True]{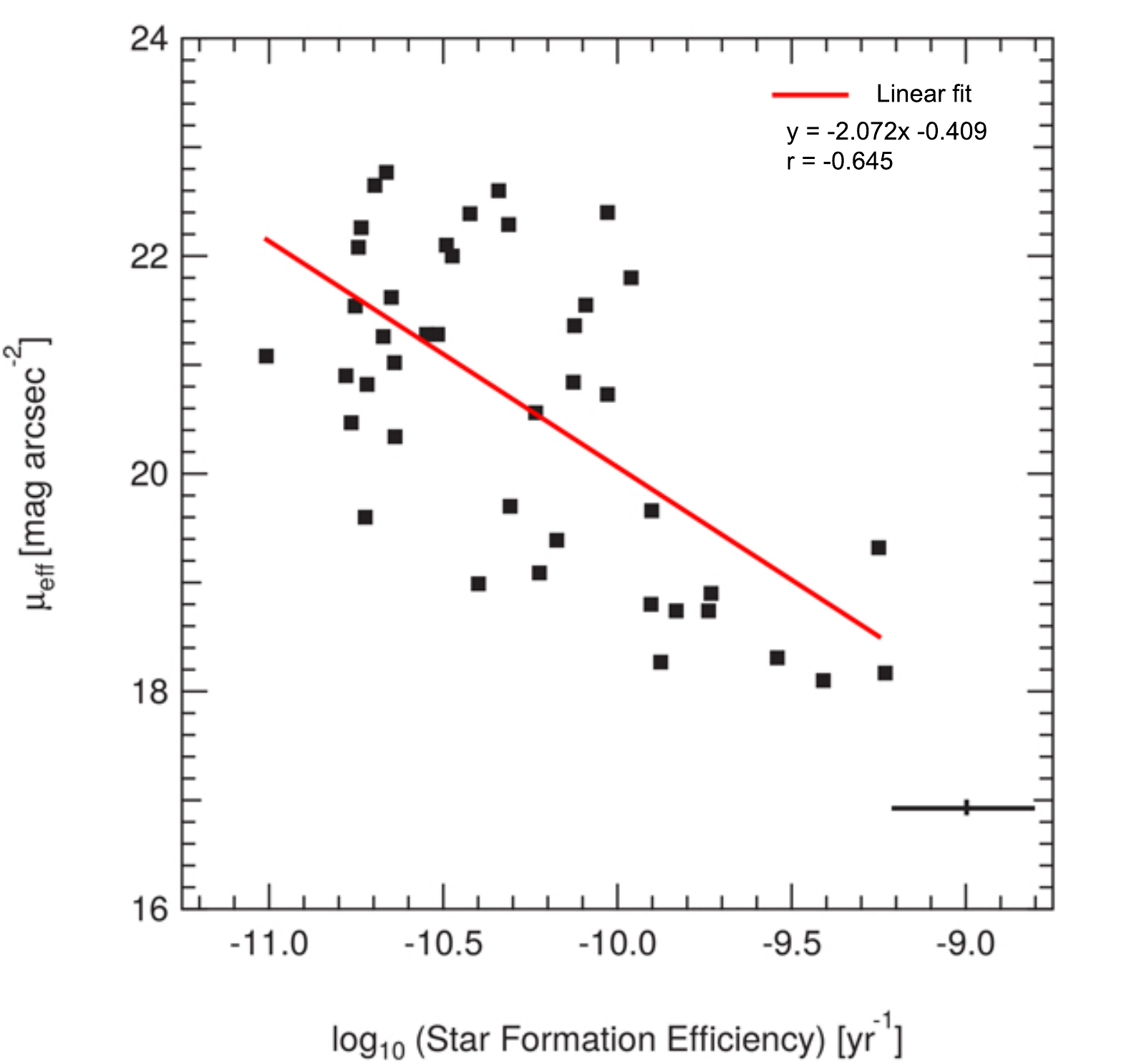}
\caption{Comparing the estimated star formation efficiency against the NIR mean surface brightness, for galaxies in the YJLK14 and KRDJ08 samples. The cross at the bottom-right shows representative $1\sigma$ errorbars. The red continuous line is a linear fit to the data.}
\label{muvSFE}
\end{figure}

In their analysis of a sample of low surface brightness (LSB) galaxies \citet{Wyder2009} suggested that a relative decrease in the molecular fraction in environments below the saturation limit, \mbox{$\Sigma_{\HIsub}$ $\approx$ 9 M$_{\odot}$ pc$^{-2}$} as a possible explanation for the decreased SFE. Following the suggestions of \cite{Blitz2006}, that the ratio of molecular-to-atomic gas in galaxies is determined by hydrostatic pressure in the ISM, which is a function of the stellar surface density, the gas surface density, and the gas-to-stellar velocity dispersion ratio, \cite{Wyder2009} argues that the LSB galaxies will invariably have a lower ISM pressure and thus lower molecular fraction. 

We explore this possibility by examining the \textit{H}-band mean effective surface brightness against the star formation efficiency for the  KRDJ08 and YJLK14 samples in Fig.~\ref{muvSFE}. 
As discussed in KRDJ08  and YJLK14, we can use the  \textit{H}-band surface brightness as a proxy for stellar surface density, $\Sigma_{\rm star}$.
The data in Fig.~\ref{muvSFE} covers six orders of magnitude in surface brightness and is well scattered with a minor trend towards higher SFE with higher surface brightness. A linear fit to the data (red continuous line in Fig.~\ref{muvSFE}) provides a correlation coefficient of $r=-0.645$. 
This suggests that the stellar densities play a minor role in setting the SFE amongst the various dwarf galaxies. 
Many of the galaxies in Fig.~\ref{muvSFE} approach surface brightnesses typical of late type spirals found in the 2MASS Large Galaxy Atlas \citep{Jarrett2003} yet have a much larger range of SFEs. In relative terms, it would appear that the lower stellar densities relative to the spiral population play a minor role in setting the SFE.

However, we note that, following \citet{Shi2011} and \citet{Wang17}, an effect of the $\Sigma_{\rm star}$ should be expected, as introduced in the extended Schmidt law. Recently \citet{Roychowdhury2017} used data from the FIGGS survey to show that low-metallicity faint dwarf galaxies also followed the extended Schmidt law. They found that the mean deviation of the FIGGS galaxies from the extended Schmidt law is 0.01~dex, with a scatter around the relation of less than half that seen in the original \citet{Shi2011} relation. A precise determination of the SFE (i.e, performing a detailed SFR study and deriving the amount of neutral gas using interferometric maps) should be obtained to explore further this issue using our galaxy sample

\section{Comparisons to models} \label{modelsec}

In the previous sections, we attempted to describe the vertical offset displayed by the dwarf galaxies in Fig.~\ref{KSlaw} in terms of variations in their disk or molecular fraction. \cite{Krumholz2012} demonstrated that in the case of the molecular star formation relation, much of the observed scatter could be reduced by normalizing the molecular gas mass per free fall time,
\begin{align}
&\Sigma_{\text{SFR}} = f_{\text{\textit{H}}_2}\epsilon_{\text{ff}}\frac{\Sigma_{\text{gas}}}{t_{\text{ff}}} \label{sflaw}\\
&t_{\text{ff}} = \sqrt{\frac{3\pi}{32G\rho}} \label{tff}
\end{align}
where $\rho$ is the density of the star-forming complex, $f_{\text{\textit{H}}_2}$ is the fraction of the molecular gas and $\epsilon_{\text{ff}}$ is a dimensionless SFE scale factor. Since in extra-galactic observations of regular spirals the resolution does not approach that of a giant molecular cloud, the mean surface gas densities are instead representative of the ISM. In the case where star formation occurs in a galaxy primarily through GMC complexes, \cite{Krumholz2012} show that the free fall time can be estimated from easily observed projected quantities,
\begin{align}
t_{\text{ff}} = \frac{\pi^{1/4}}{\sqrt{8}}\frac{\sigma}{G\left(\Sigma^3_{\text{GMC}}\Sigma_{\text{gal}}\right)^{1/4}} \label{tffgmc}
\end{align}
where $\Sigma_{\text{GMC}}$ is the mean {gas} surface density of GMCs value adopted from local observations, $\sigma$ is the local velocity dispersion of the gas and $\Sigma_{\text{gal}}$ is the average {gas} surface density in the region of the galaxy where the GMCs form. Since variations in disk scale height should result in relative changes of the volume density of the gas and therefore the free fall time, for a given mean value of the gas velocity dispersion within the star-forming disk, the free fall time could be estimated. Applying Equation \ref{sflaw}, having now accounted for projection effects, any remaining vertical offset between dwarf and spiral galaxies could be more precisely attributed to variations in $f_{H_2}$, the molecular fraction of the gas.

As demonstrated by \cite{Leroy2013} and \cite{Shetty2014}, any star formation relation is ultimately a multidimensional problem and potentially even a multivalued function of $\Sigma_{\text{gas}}$. Using observations to constrain the inputs, comparisons to models with varying underlying theoretical assumptions could provide greater insight into the processes which regulate star formation. Indeed, the correlation between atomic gas and the SFR surface densities in \HI-dominated regimes, as demonstrated in this study and others \citep{Bigiel2010,Bolatto2011,Roychowdhury2014,Roychowdhury2015}, has been the subject of interest in recent models \citep{Ostriker2010,Krumholz2013}, whose underlying theoretical assumptions differ. 

In the \cite{Ostriker2010} model (hereafter OML), the ISM is comprised of warm and cold diffuse gas and a gravitationally bound phase. Star formation occurs in the gravitionally bound component without explicit treatment of its chemical state, at a rate corresponding to a fiducial choice in gas depletion timescale of 2~Gyr. The ISM satisfies thermal and vertical hydrostatic equilibrium whereby the UV heating (proportional to the star formation rate) can be balanced against the midplane density. The transition from \HI-to H$_2$-dominated gas, and the effect gas metallicity has on it in the OML10 model, is not treated explicitly but manifests itself instead where the composition of the ISM transitions from mostly diffuse to mostly gravitationally bound. This transition happens smoothly, and will be an important factor when comparing models later. Alternatively, the \citet[][hereafter, KMT+]{Krumholz2013} model, a low surface density extension to \cite[][hereafter KMT]{Krumholz2012}, explicitly determines the fraction of gas in the molecular phase which corresponds to the gas phase that is eligible to form stars. The fraction of the gas that is H$_2$ is calculated from the gas metallicity through dust shielding  and through the photo-dissociation rate, as described in \citep{Krumholz2008,Krumholz2009a}. Perhaps motivated by a desire to reproduce recent resolved observations of \HI-dominated galaxies \citep{Bigiel2010} and the SMC \citep{Bolatto2011}, \cite{Krumholz2013} successfully developed a model which predicts a floor in the star formation rate surface density. They argue that the interstellar radiation field (ISRF) intensity is not sufficient enough to satisfy hydrostatic equilibrium in the ISM of \HI-dominated galaxies. By considering hydrostatic balance alone, a 'floor' value on the density of the cold atomic phase can be determined, which can then be equated to a molecular fraction and star formation rate.

\begin{figure}
\center
\includegraphics[trim = 1mm 0mm 0mm 0mm, scale=0.5, clip=True]{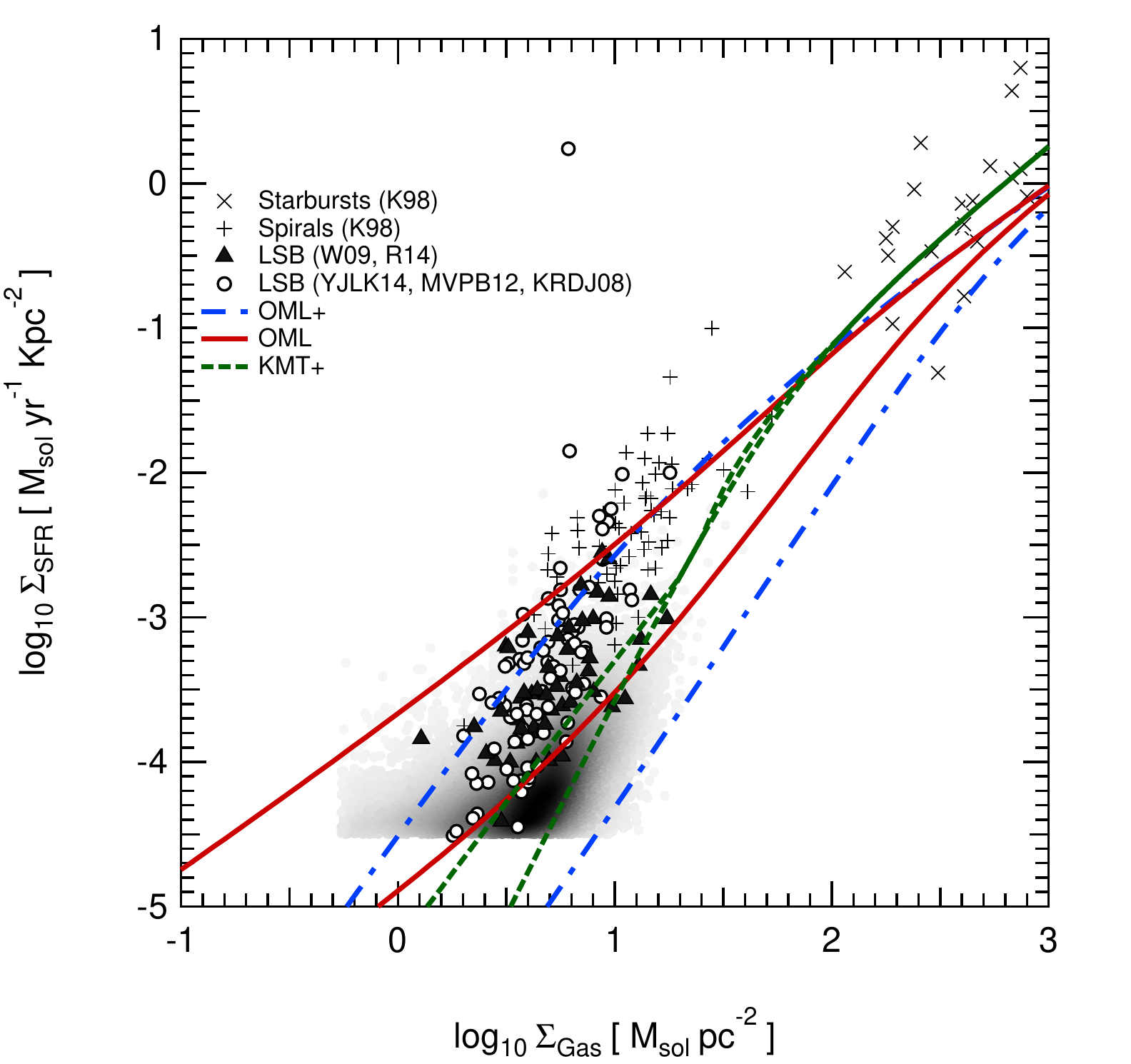}
\caption{The log star formation rate versus total gas surface densities. Dwarf galaxies are assumed to have a negligible molecular fraction and so their total gas surface density is simply the atomic gas corrected for the presence of helium using a factor of 1.34. The K98 galaxies have been corrected for the presence of helium contrary to the original work. Filled triangles indicate disk averaged quantities for the galaxies whose column densities were measured directly from the \protect\cite{Roychowdhury2014,Wyder2009} studies. Open circles indicate galaxies whose total gas surface densities have been estimated using global \HI properties as described in the text. The greyscale smoothed distribution represents the outer disk data of \HI-dominated dwarf and spirals galaxies from \protect\cite{Bigiel2010} (shown in Fig.~\ref{KSlaw} as stars with errorbars). The models are indicated by the lines in the legend and are described in detail in the text.}
\label{modelcomp}
\end{figure}

Figure \ref{modelcomp} compares the KMT+, OML and the \citet[][hereafter, OML+]{Bolatto2011} (modification of the original OML by introducing an extra metallicity dependence). These models have been computed with the following parameters. Based on the discussions of \cite{Ostriker2010}, and \cite{Krumholz2013} we adopted a ratio of the mass-weighted mean thermal velocity dispersion, where if significant warm neutral medium is present, is approximately equal to the fraction of diffuse warm gas, \mbox{$\tilde{f}_w \approx f_{w}$ = 0.5.} Variations in this quantity are degenerate with the midplane density parameter, $\rho_{\text{sd}}$, for which we have adopted values in the range of 0.3-0.003 $M_{\odot}$ pc$^{-3}$ in line with the ranges in other studies \citep[e.g.][]{Roychowdhury2015,Roychowdhury2014,Krumholz2013}. For the OML and OML+ models, values of the observed total velocity dispersion divided by the mean thermal value are ranged from two to 10 as per their discussions to produce the widest range in model tracks. For the KMT+ model, a clumping factor of $f_c$ = 5 is adopted, which is the expected reduction factor in the mean surface density of cloud complexes when averaged over the ISM on kpc scales. All models are computed with a stellar metallicity of Z = 0.1~Z$_{\odot}$.

Figure \ref{modelcomp} illustrates the success with which these models describe various datasets. The \HI-dominated outer disk data from \cite{Bigiel2010} is well described by the OML+ and KMT+ models. 
However, \citet{Roychowdhury2015} observations matched better with the predictions of the OML model than those provided by the KMT model for galaxies of all types.
The physical implications of the introduced metallicity dependence are unclear  for the OML+ model \citep[see section 4.3 of][]{Krumholz2013}. In the KMT+ model, this metallicity dependent transition arises naturally from self-shielding of the molecular gas, however since the contrast between the self-shielded H$_2$ and transparent gas is very sharp, the KMT tracks converge faster than the OML and OML+ models (in these models the metallicity is degenerate with the midplane density parameter). The KMT+ model appears to be the most successful model when providing a 'floor' on the SFE and describing the transition from \HI-dominated to H$_2$-dominated gas. 

\begin{figure}
\center
\includegraphics[trim = 1mm 0mm 0mm 0mm, scale=0.5, clip=True]{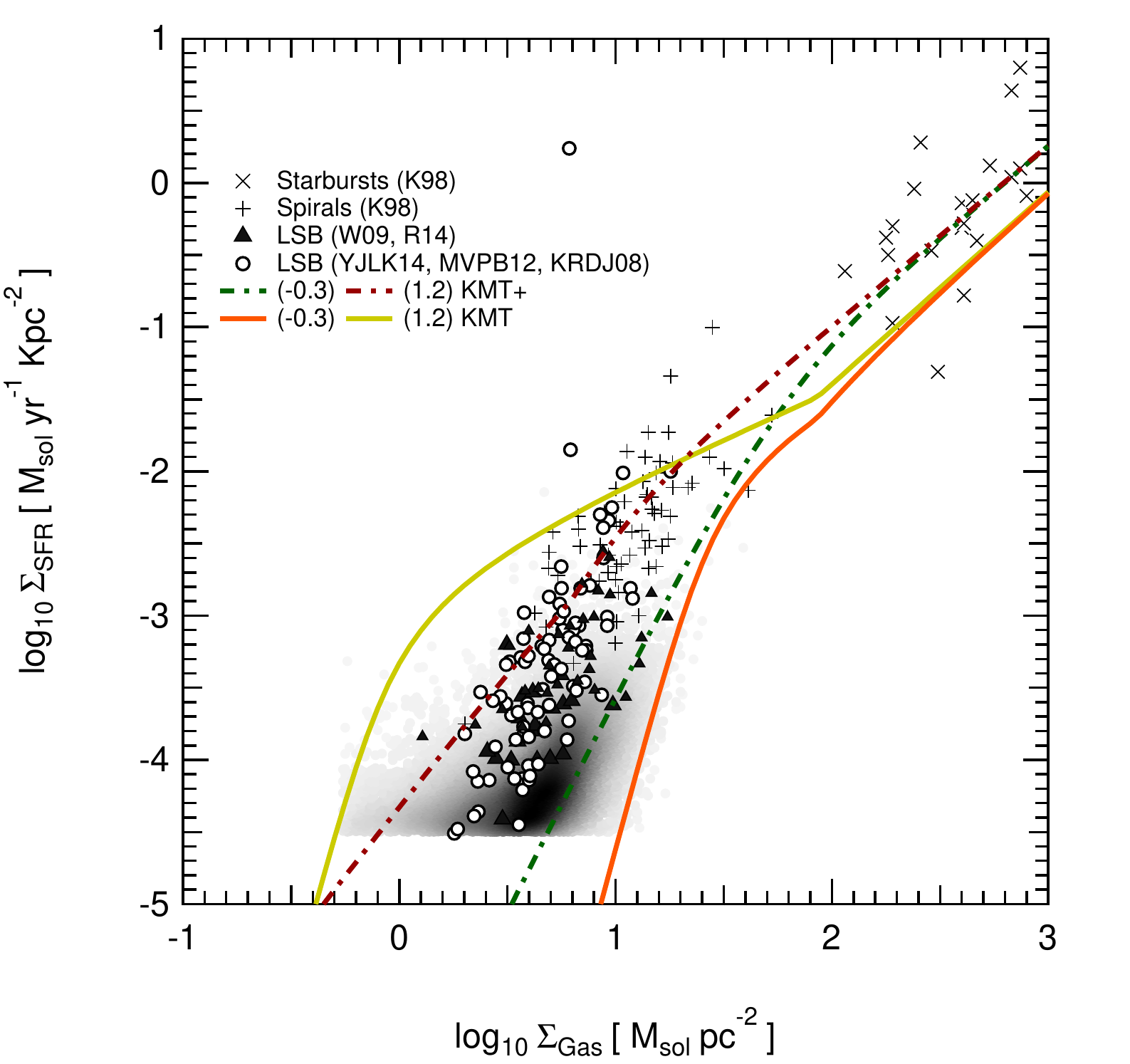}
\caption{As in Fig.~\ref{modelcomp} but exclusively for the KMT and KMT+ models over a range of clumping factors, $-$0.3 $<$ $\log_{10}(f_cZ_*)$ $<$ 1.2. The agreement between the KMT model and the data is good but requires an unreasonably large variation in clumping factors (see text for further details).}
\label{clumpingfactor}
\end{figure}

Over the range of adopted fiducial parameters however, the KMT+ model can not adequately describe the disk-averaged dwarf and low-mass late type galaxy measurements as shown in Fig.~\ref{modelcomp} as open circles and filled triangles respectively. 
We consider that this is unlikely the result of a metallicity effect. Following the stellar mass-metallicity relation presented by \cite{Kirby2013}, galaxies with stellar masses less than 10$^{9}$ \msun  are not expected to have stellar metallicities higher than \mbox{Z = 0.1 Z$_{\odot}$} (see their Fig.~9). We clarify we are referring to stellar metallicities, galaxies with stellar masses less than 10$^{9}$ \msun will typically have gas-phase metallicities in the range of \mbox{Z = 0.08 -- 0.3 Z$_{\odot}$}  \citep[e.g.][]{LS10}.
The OML and OML+ models are not particularly sensitive to metallicity over the range of 0.01$<$ Z$_*$ $<$ 0.1, and these variations are degenerate with stellar density. Similarly in the KMT+ model, metallicity variations do not change the overall picture, rather metallicity determines the location of the transition from \HI-to-H$_2$ dominated gas. 

The implication is that in the outer disks of spirals and dwarf galaxies, hydrostatic balance alone is sufficient to describe the star formation rates, but mean values of the overall disks of these low-mass galaxies suggest that thermal pressure provides an important contribution and is required to explain the full range of the data. This is consistent with the conclusions of \cite{Elmegreen2015}, who found that models considering thresholds in the presence of H$_2$ did not agree with the smooth trends observed for the disks of the dIrr sample.


It is worth noting that 2 orders of magnitude in midplane density is quite generous, and although in general we do not have a handle on this quantity, 0.3 $M_{\odot}$ pc$^{-3}$ is already a factor of ten times larger than values that have been inferred in galaxies \citep[e.g.][]{Bruzzese2015}. Restricting this value further suggests that the OML model under-predicts the observed star formation rate densities. Alternatively, the original KMT model \citep{Krumholz2009} can describe the full range of the data if one relaxes the restriction on the clumping factor $f_c$. Figure \ref{clumpingfactor} compares the KMT and KMT+ models with a range of $-$0.3 $<$ $\log_{10}(f_cZ_*)$ $<$ 1.2, and in the case of the KMT+ model, a range of midplane densities an order of magnitude smaller than in Fig.~\ref{modelcomp} \mbox{(0.03-0.003 $M_{\odot}$ pc$^{-3}$)} was adopted. The original KMT model adequately describes the full range of the data with these model parameters. However while perhaps appropriate when describing variation in spiral galaxies, a $\log_{10}(f_cZ_*)$ value of 1.2 for a metallicity Z$_*$ = 0.1 in units of solar metallicity suggests clumping factors an order of magnitude higher than that fiducial value of five adopted here and in tension with observations \citep{Leroy2013a}.

Variations in the underlying IMF and non-constant SFHs leading to systematic variations in the SFR estimates may explain this discrepancy. Although a full treatment of this topic is outside the scope of this work, we redirect the reader to \citet{Boquien2014}, \cite{Weisz2012} and \cite{Fumagalli2011}. Recent observations with the STARBIRDS sample of galaxies by \cite{McQuinn2015} suggest an empirical calibration of the FUV-based SFR relation for dwarf galaxies, which is $\sim$53\% more than \cite{Hao2011} which would increase the SFRs by a factor of 1.5 and further worsens agreement between the models and data presented here.

\section{Summary} \label{summary}
We have provided a method for estimating the surface density of the atomic gas from global \HI parameters which are widely available from \HI surveys, and precision geometric parameters available from stellar photometry. We test this method using two control samples and find the approximation overestimates the actual surface gas density in \HI-dominated galaxies to within a factor of 0.5 dex. We apply this method to a sample of 147 galaxies drawn from modern NIR stellar photometric surveys cross correlated with available \textit{FUV} and 24$\mu$m fluxes to estimate the star formation rate surface density.

With this sample we confirm a strict correlation between the \HI surface gas density and the SFR surface density with a sample of dwarf galaxies an order of magnitude larger than previous works. We find that the SFE of \HI-dominated gas is offset from that in spirals. With considerations for variations of disk scale heights, we infer from relative comparison to molecular star formation relations that the mean molecular fractions in low-mass late types fall within 5-15\%. 
Our analysis suggests that the stellar densities play a minor role in setting the SFE amongst the various dwarf galaxies, although following recent work we consider
that using accurate, interferometric \HI data we might recover the proposed extended Schmidt law which relates $\Sigma_{\rm SFR}$, $\Sigma_{\rm gas}$ and $\Sigma_{\rm star}$.


Finally we have compared our data and others from the literature to available models. We show that no single model can describe the full extent of the data. However these models together suggest that feedback is of variable importance depending on the density regime or galaxy environment, and that the transition from \HI to H$_2$-dominated is a metallicity dependent process. Mean values of $\Sigma_{\text{SFR}}$ and $\Sigma_{\text{gas}}$ suggest that globally in these \HI-dominated galaxies, thermal pressure is an important regulatory process.

\section*{Acknowledgments}
CL is funded by a Discovery Early Career Researcher Award (DE150100618). TY would like to thank the Research School of Astronomy and Astrophysics, CSIRO Astronomy and Space Science division and the Australian Astronomical Observatory through their support with the Research Training Scheme, ATNF Graduate program and the Trevor Burgess Scholarship respectively. HJ acknowledges the support of the Australian Research Council through Discovery Projects DP120100475 and DP 150100862. The Authors would like to extend their gratitude to \cite{Roychowdhury2014} for kindly providing a copy of their data. The authors also thank the anonymous referee for their very constructive comments that have helped to improve the quality of this paper. Parts of this research were conducted by the Australian Research Council Centre of Excellence for All-sky Astrophysics (CAASTRO), through project number CE110001020. Parts of this research were supported by the Australian Research Council Centre of Excellence for All Sky Astrophysics in 3 Dimensions (ASTRO 3D), through project number CE170100013.

\section*{Appendix}

We include here the table listing the externally obtained properties required to compute the surface densities for the KRDJ08 and YJLK14 NIR samples (Table~\ref{sampleq_appendix}) and the table listing the computed quantities for the NIR catalogue galaxies (Table~\ref{sampled_appendix}). 

\label{lastpage}

\footnotesize
\bibliographystyle{mn2e}
\bibliography{bibclean}

\clearpage

\begin{deluxetable}{lcccccccccccccc} 
\tabletypesize{\scriptsize}
\tablecaption{Table listing the externally obtained properties required to compute the surface densities for the KRDJ08 and YJLK14 NIR samples. 
The Table is arranged as follows: Column 1 - Galaxy name; Columns 2 and 3 - RA and DEC; Columns 4 - Morphology, using the classification scheme by \protect\cite{DeVaucouleurs1991}; Columns 5 and 6 - the radial distance in Mpc and the derivation method respectively. Note that obtained values have been rounded to the nearest decimal place since distance derivation methods are typically no more accurate than 10\%; Column 7 - FUV flux, Column 8 - the 24$\mu$m luminosity; Column 9 - the \HI flux; Column 10 - the \textit{B}-band galactic extinction;  Column 11 - the total observed luminosity in the \textit{B}-band;  Column 12 - \textit{B-H} colour corrected for extinction (or the \textit{B-Ks} data for the MVPB12 sample). Quantities are carefully cited in the caption for the KRDJ08 and YJLK14 samples. For the MVPB12 sample, their values have been taken as is and when available (such as distance modulus, \HI fluxes, extinctions etc). The authors did not, however compile a list of \textit{B}-band magnitudes, which we require in our computation of $R_{26.5,eq}$. Instead, we obtained them using the \protect\cite{DeVaucouleurs1991} RC3 catalogue. Similarly, the quantities calculated for \protect\citet{Kennicutt1998} and \protect\citet{Tully1988} cross correlated galaxies are not presented but these data can be made available upon request.
References: \textbf{\textit{Distance}} (a) \protect\citet{Karachentsev2002}, (b) \protect\citet{Karachentsev2003}, (c) \protect\citet{Karachentsev2006}, (d) \protect\citet{Karachentsev2007}, (e) \protect\citet{Karachentsev2013}, (f) \citet{Tully2006}, (g) \citet{Tully2008}, (h) \protect\citet{Karachentsev2003a}, (i) \protect\citet{Karachentsev2002a},  (j) \protect\citet{Karachentsev2000}, (k) \protect\citet{Seth2005}, (l) \protect\citet{Roychowdhury2012}, (m) \protect\citet{Cannon2003} , (n) \protect\citet{Grise2008}, (o) \protect\citet{Tosi2001}, (p) \protect\citet{Tonry2001}, (q) \protect\citet{Dalcanton2009}; \textbf{\textit{UV-band}} (a) \textit{GALEX}, (b) \protect\citet{Lee2011}; \textbf{\textit{B-band}} (a) \protect\citet{DeVaucouleurs1991}, (b) \protect\citet{Lauberts1989}, (c) \protect\citet{Karachentsev2013}, (d) \protect\citet{Metcalfe1994}, (e) \protect\citet{Parodi2002}, (f) \protect\citet{Jerjen2000}, (g) \protect\citet{Warren2007}, (h) \citet{Karachentsev2004}, (i) \protect\citet{Warren2006}, (j) \protect\citet{Kouwenhoven2007}, (k) \protect\citet{Roychowdhury2012}; \textbf{\textit{H}} \textsc{i} \textbf{Flux}, (a) \protect\citet{Doyle2005}, (b) \protect\citet{Begum2008}, (c) \protect\citet{Bouchard2005}. For references pertaining to the MVPB12 data, the reader should consult their study.\label{sampleq_appendix}}
\tablewidth{0pt}
\tablehead{
\colhead{Galaxy Name} &
\colhead{RA} &
\colhead{DEC} &
\colhead{Type} &
\colhead{D} &
\colhead{Method} &
\colhead{m$_{FUV}$} &
\colhead{L(24$\mu$m)} &
\colhead{F\HI} &
\colhead{$A_B$} &
\colhead{$m_B$} &
\colhead{$B-H$} \\

	& 	(J2000)	& 	(J2000)	& 		& 	[Mpc]	&		&	[mag]	&	[erg s$^{-1}$]	&	 [Jy km s$^{-1}$]	&	[mag]	&	[mag]	&	[mag]	&	\\
(1)	& 	(2)	& 	(3)	& 	(4)	& 	(5)	&	(6)	&	(7)	&	(8)	&	(9)	&	(10)	&	(11)	&	(12)	
}

\startdata

YJLK14						\\	
\noalign{\smallskip}
																	
\hline																								
AM1321-304	&	13 24 36.0	&	$-$30 58 20	&	10	&	4.6$^{a}$	&	TRGB	&	18.8$^{a}$	&	-	&	1.6$^{b}$	&	0.25	&	16.7$^{d}$	&	3.2	&	\\
CEN06	&	13 05 02.1	&	$-$40 04 58	&	10	&	5.8$^{d}$	&	TRGB	&	-	&	-	&	5.1$^{a}$	&	0.37	&	17.7$^{f}$	&	2.9	&	\\
ESO149-G003	&	23 52 02.8	&	$-$52 34 39	&	10	&	5.9$^{a}$	&	TF	&	15.7$^{b}$	&	38.67	&	6.9$^{h}$	&	0.05	&	15.1$^{b}$	&	1.9	&	\\
ESO199-G007	&	02 58 04.1	&	-49 22 57	&	10	&	6.6	&	h	&	17.9$^{a}$	&	-	&	2.1$^{a}$	&	0.08	&	16.4$^{b}$	&	2.1	&	\\
ESO222-G010	&	14 35 03.0	&	-49 25 18	&	10	&	5.8$^{e}$	&	TF	&	-	&	-	&	7.0$^{a}$	&	0.8	&	16.3$^{h}$	&	3.2	&	\\
ESO223-G009	&	15 01 08.5	&	-48 17 33	&	10	&	6.5$^{d}$	&	TRGB	&	-	&	-	&	101.3$^{a}$	&	0.94	&	13.8$^{h}$	&	1.9	&	\\
ESO252-IG001	&	04 56 58.7	&	-42 48 14	&	10	&	7.2$^{e}$	&	TF	&	-	&	-	&	10.9$^{a}$	&	0.05	&	14.4$^{a}$	&	1.2	&	\\
ESO269-G058	&	13 10 32.9	&	-46 59 27	&	9	&	3.8$^{d}$	&	TRGB	&	17.2$^{a}$	&	-	&	5.3$^{a}$	&	0.39	&	13.3$^{a}$	&	3.2	&	\\
ESO272-G025	&	14 43 25.5	&	-44 42 19	&	10	&	5.9$^{a}$	&	h	&	-	&	-	&	6.9$^{a}$	&	0.58	&	14.8$^{b}$	&	2.6	&	\\

\noalign{\smallskip}
\hline
\enddata
\end{deluxetable}


\begin{deluxetable}{lcc ccc cc} 
\tablecaption{Table listing the computed quantities for the NIR catalogue galaxies, and is arranged as follows : Column 1 - Name, Column 2 - $L(FUV)_{\text{corr}}$, Column 3 - $\dot{M}_*$, Column 4 - {$M_{\text{\HI}}$}, Column 5 - $R_{26.5,eq}$, Column 6 - $\Sigma_{\HIsub}$, Column 7 - $\Sigma_{\text{SFR}}$.\label{sampled_appendix}}
\tablewidth{0pt}
\tablehead{
\colhead{Galaxy Name} &
\colhead{$L(FUV)_{\text{corr}}$} &
\colhead{$\dot{M}_*$} &
\colhead{$M_{\HIsub}$} &
\colhead{$R_{26,eq}$} &
\colhead{$\Sigma_{\HIsub}$} &
\colhead{$\Sigma_{SFR}$} \\
	&	[erg	s$^{-1}$]	&	[$M_{\odot}$	yr$^{-1}$]	&	[$M_{\odot}$]	&	[pc]	&	$\log[M_{\odot}$	$pc^{-2}]$	&	$\log[M_{\odot}$	$yr^{-1}$	$Kpc^{-2}]$

}

\startdata

\noalign{\smallskip}
YJLK14	\\																							
\noalign{\smallskip}
\hline																								
AM0106-382	&	-	&	-	&	-	&	1222	&	&	\\													
CEN06	&	-	&	-	&	7.6	&	761	&	1.16	&	-	\\											
ESO149-G003	&	41.24	&	-1.90	&	7.8	&	1761	&	0.22	&	-3.44	\\											
ESO199-G007	&	40.48	&	-2.67	&	7.3	&	1476	&	0.07	&	-3.93	\\											
ESO222-G010	&	-	&	-	&	7.7	&	1708	&	0.66	&	-	\\											
ESO223-G009	&	-	&	-	&	9.0	&	3132	&	1.37	&	-	\\											
ESO252-IG001	&	-	&	-	&	8.1	&	1839	&	0.84	&	-	\\											
ESO269-G058	&	40.67	&	-2.48	&	7.3	&	1801	&	0.08	&	-3.66	\\											
ESO272-G025	&	-	&	-	&	7.7	&	1873	&	0.16	&	-	\\											
\noalign{\smallskip}
\hline
\enddata
\end{deluxetable}

\end{document}


\begin{deluxetable}{lcccccccccccccc} 
\tabletypesize{\scriptsize}
\tablecaption{Table listing the externally obtained properties required to compute the surface densities for the KRDJ08 and YJLK14 NIR samples. 
The Table is arranged as follows: Column 1 - Galaxy name; Columns 2 and 3 - RA and DEC; Columns 4 - Morphology, using the classification scheme by \protect\cite{DeVaucouleurs1991}; Columns 5 and 6 - the radial distance in Mpc and the derivation method respectively. Note that obtained values have been rounded to the nearest decimal place since distance derivation methods are typically no more accurate than 10\%; Column 7 - FUV flux, Column 8 - the 24$\mu$m luminosity; Column 9 - the \HI flux; Column 10 - the \textit{B}-band galactic extinction;  Column 11 - the total observed luminosity in the \textit{B}-band;  Column 12 - \textit{B-H} colour corrected for extinction (or the \textit{B-Ks} data for the MVPB12 sample). Quantities are carefully cited in the caption for the KRDJ08 and YJLK14 samples. For the MVPB12 sample, their values have been taken as is and when available (such as distance modulus, \HI fluxes, extinctions etc). The authors did not, however compile a list of \textit{B}-band magnitudes, which we require in our computation of $R_{26.5,eq}$. Instead, we obtained them using the \protect\cite{DeVaucouleurs1991} RC3 catalogue. Similarly, the quantities calculated for \protect\citet{Kennicutt1998} and \protect\citet{Tully1988} cross correlated galaxies are not presented but these data can be made available upon request.
References: \textbf{\textit{Distance}} (a) \protect\citet{Karachentsev2002}, (b) \protect\citet{Karachentsev2003}, (c) \protect\citet{Karachentsev2006}, (d) \protect\citet{Karachentsev2007}, (e) \protect\citet{Karachentsev2013}, (f) \citet{Tully2006}, (g) \citet{Tully2008}, (h) \protect\citet{Karachentsev2003a}, (i) \protect\citet{Karachentsev2002a},  (j) \protect\citet{Karachentsev2000}, (k) \protect\citet{Seth2005}, (l) \protect\citet{Roychowdhury2012}, (m) \protect\citet{Cannon2003} , (n) \protect\citet{Grise2008}, (o) \protect\citet{Tosi2001}, (p) \protect\citet{Tonry2001}, (q) \protect\citet{Dalcanton2009}; \textbf{\textit{UV-band}} (a) \textit{GALEX}, (b) \protect\citet{Lee2011}; \textbf{\textit{B-band}} (a) \protect\citet{DeVaucouleurs1991}, (b) \protect\citet{Lauberts1989}, (c) \protect\citet{Karachentsev2013}, (d) \protect\citet{Metcalfe1994}, (e) \protect\citet{Parodi2002}, (f) \protect\citet{Jerjen2000}, (g) \protect\citet{Warren2007}, (h) \citet{Karachentsev2004}, (i) \protect\citet{Warren2006}, (j) \protect\citet{Kouwenhoven2007}, (k) \protect\citet{Roychowdhury2012}; \textbf{\textit{H}} \textsc{i} \textbf{Flux}, (a) \protect\citet{Doyle2005}, (b) \protect\citet{Begum2008}, (c) \protect\citet{Bouchard2005}. For references pertaining to the MVPB12 data, the reader should consult their study.\label{sampleq_appendix}}
\tablewidth{0pt}
\tablehead{
\colhead{Galaxy Name} &
\colhead{RA} &
\colhead{DEC} &
\colhead{Type} &
\colhead{D} &
\colhead{Method} &
\colhead{m$_{FUV}$} &
\colhead{L(24$\mu$m)} &
\colhead{F\HI} &
\colhead{$A_B$} &
\colhead{$m_B$} &
\colhead{$B-H$} \\

	& 	(J2000)	& 	(J2000)	& 		& 	[Mpc]	&		&	[mag]	&	[erg s$^{-1}$]	&	 [Jy km s$^{-1}$]	&	[mag]	&	[mag]	&	[mag]	&	\\
(1)	& 	(2)	& 	(3)	& 	(4)	& 	(5)	&	(6)	&	(7)	&	(8)	&	(9)	&	(10)	&	(11)	&	(12)	
}

\startdata

YJLK14						\\	
\noalign{\smallskip}
																	
\hline																								
AM1321-304	&	13 24 36.0	&	$-$30 58 20	&	10	&	4.6$^{a}$	&	TRGB	&	18.8$^{a}$	&	-	&	1.6$^{b}$	&	0.25	&	16.7$^{d}$	&	3.2	&	\\
CEN06	&	13 05 02.1	&	$-$40 04 58	&	10	&	5.8$^{d}$	&	TRGB	&	-	&	-	&	5.1$^{a}$	&	0.37	&	17.7$^{f}$	&	2.9	&	\\
ESO149-G003	&	23 52 02.8	&	$-$52 34 39	&	10	&	5.9$^{a}$	&	TF	&	15.7$^{b}$	&	38.67	&	6.9$^{h}$	&	0.05	&	15.1$^{b}$	&	1.9	&	\\
ESO199-G007	&	02 58 04.1	&	-49 22 57	&	10	&	6.6	&	h	&	17.9$^{a}$	&	-	&	2.1$^{a}$	&	0.08	&	16.4$^{b}$	&	2.1	&	\\
ESO222-G010	&	14 35 03.0	&	-49 25 18	&	10	&	5.8$^{e}$	&	TF	&	-	&	-	&	7.0$^{a}$	&	0.8	&	16.3$^{h}$	&	3.2	&	\\
ESO223-G009	&	15 01 08.5	&	-48 17 33	&	10	&	6.5$^{d}$	&	TRGB	&	-	&	-	&	101.3$^{a}$	&	0.94	&	13.8$^{h}$	&	1.9	&	\\
ESO252-IG001	&	04 56 58.7	&	-42 48 14	&	10	&	7.2$^{e}$	&	TF	&	-	&	-	&	10.9$^{a}$	&	0.05	&	14.4$^{a}$	&	1.2	&	\\
ESO269-G058	&	13 10 32.9	&	-46 59 27	&	9	&	3.8$^{d}$	&	TRGB	&	17.2$^{a}$	&	-	&	5.3$^{a}$	&	0.39	&	13.3$^{a}$	&	3.2	&	\\
ESO272-G025	&	14 43 25.5	&	-44 42 19	&	10	&	5.9$^{a}$	&	h	&	-	&	-	&	6.9$^{a}$	&	0.58	&	14.8$^{b}$	&	2.6	&	\\
ESO274-G001	&	15 14 13.5	&	-46 48 45	&	7	&	3.1$^{d}$	&	TRGB	&	-	&	-	&	120.2$^{a}$	&	0.93	&	11.7$^{a}$	&	1.3	&	\\
ESO318-G013	&	10 47 41.9	&	-38 51 15	&	8	&	6.5$^{e}$	&	TF	&	16.9$^{a}$	&	-	&	8.6$^{a}$	&	0.27	&	15.0$^{h}$	&	2.2	&	\\
ESO320-G014	&	11 37 53.4	&	-39 13 14	&	10	&	6.1$^{i}$	&	TRGB	&	19.0$^{a}$	&	-	&	2.5$^{a}$	&	0.52	&	15.9$^{b}$	&	2.2	&	\\
ESO321-G014	&	12 13 49.6	&	-38 13 53	&	10	&	3.2$^{a}$	&	TRGB	&	16.4$^{b}$	&	37.81	&	6.4$^{a}$	&	0.34	&	15.2$^{b}$	&	2.2	&	\\
ESO325-G011	&	13 45 00.8	&	-41 51 32	&	10	&	3.4$^{a}$	&	TRGB	&	-	&	-	&	26.6$^{a}$	&	0.32	&	14.0$^{b}$	&	1.4	&	\\
ESO376-G016	&	10 43 27.1	&	-37 02 33	&	10	&	7.1$^{e}$	&	TF	&	-	&	-	&	10.3$^{a}$	&	0.21	&	15.5$^{a}$	&	0.9	&	\\
ESO379-G007	&	11 54 43.0	&	-33 33 29	&	10	&	5.2$^{a}$	&	TRGB	&	-	&	-	&	5.2$^{a}$	&	0.27	&	16.6$^{a}$	&	2.2	&	\\
ESO379-G024	&	12 04 56.7	&	-35 44 35	&	10	&	4.9	&	h	&	-	&	-	&	3.4$^{a}$	&	0.28	&	16.6$^{a}$	&	2.6	&	\\
ESO410-G005	&	00 15 31.4	&	-32 10 48	&	10	&	1.9$^{j}$	&	TRGB	&	18.3$^{b}$	&	38.40	&	0.8$^{c}$	&	0.05	&	14.9$^{e}$	&	2.5	&	\\
ESO444-G084	&	13 37 20.2	&	-28 02 46	&	10	&	4.6$^{a}$	&	TRGB	&	16.0$^{b}$	&	38.90	&	21.1$^{a}$	&	0.25	&	15.1$^{h}$	&	1.6	&	\\
ESO489-G?056	&	06 26 17.0	&	-26 15 56	&	10	&	5.0$^{b}$	&	TRGB	&	-	&	-	&	2.4$^{a}$	&	0.24	&	15.7$^{e}$	&	1.8	&	\\
IC4247	&	13 26 44.4	&	-30 21 45	&	10	&	4.97$^{d}$	&	TRGB	&	16.0$^{b}$	&	39.19	&	-	&	0.23	&	14.4$^{b}$	&	2.4	&	\\
IC4316	&	13 40 18.1	&	-28 53 40	&	10	&	4.4$^{a}$	&	TRGB	&	16.1$^{b}$	&	-	&	2.2$^{b}$	&	0.2	&	14.6$^{h}$	&	2.7	&	\\
NGC0625	&	01 35 05.0	&	-41 26 11	&	8	&	3.89$^{m}$	&	TRGB	&	13.8$^{b}$	&	41.30	&	30.9$^{a}$	&	0.06	&	11.7$^{a}$	&	2.7	&	\\
NGC2188	&	06 10 09.5	&	-34 06 22	&	8	&	7.4$^{e}$	&	TF	&	-	&	-	&	32.5$^{a}$	&	0.12	&	12.1$^{a}$	&	2.4	&	\\
NGC5264	&	13 41 37.0	&	-29 54 50	&	8	&	4.53$^{i}$	&	TRGB	&	14.9$^{b}$	&	40.18	&	12.8$^{a}$	&	0.19	&	12.6$^{a}$	&	2.1	&	\\
NGC5408	&	14 03 21.5	&	-41 22 35	&	9	&	4.81$^{i}$	&	TRGB	&	-	&	-	&	61.5$^{a}$	&	0.25	&	12.2$^{a}$	&	1.2	&	\\
UGCA365	&	13 36 30.8	&	-29 14 11	&	10	&	5.25$^{d}$	&	TRGB	&	18.0$^{a}$	&	38.74	&	1.2$^{a}$	&	0.23	&	15.5$^{b}$	&	2.3	&	\\
\hline																								
\noalign{\smallskip}
KJRD08					\\	
\noalign{\smallskip}
																		
\hline																								
AM0106-382	&	01 08 22.0	&	$-$38 12 33	&	10	&	8.2$^{e}$	&	TF	&	-	&	-	&	-	&	0.05	&	16.6$^{c}$	&	2.8	&	\\
AM0319-662	&	03 21 02.4	&	$-$66 19 09	&	10	&	4.0$^{b}$	&	TRGB	&	19.8$^{a}$	&	-	&	-	&	0.3	&	16.5$^{c}$	&	2.2	&	\\
AM0333-611	&	03 34 15.3	&	$-$61 05 47	&	10	&	14.5	&	h	&	-	&	-	&	5.3$^{a}$	&	0.12	&	-	&	-	&	\\
AM0521-343	&	05 23 23.72	&	-34 34 29	&	10	&	11.8	&	h	&	-	&	-	&	4.9$^{a}$	&	0.1	&	-	&	-	&	\\
AM0737-691	&	07 37 12.7	&	-69 20 38	&	10	&	17.8	&	h	&	20.6$^{a}$	&	-	&	-	&	0.74	&	-	&	-	&	\\
Argo	&	07 05 17.1	&	-58 31 14	&	10	&	4.9$^{b}$	&	TRGB	&	-	&	-	&	34.8$^{a}$	&	0.42	&	15.0$^{e}$	&	1.9	&	\\
DDO210	&	20 46 51.8	&	-12 50 53	&	10	&	1.0$^{a}$	&	TRGB	&	16.5$^{b}$	&	37.81	&	11.2$^{a}$	&	0.18	&	14.0$^{a}$	&	1.5	&	\\
ESO006-G001	&	08 19 23.3	&	-85 08 44	&	9	&	6.7	&	h	&	-	&	-	&	-	&	0.67	&	15.1$^{f}$	&	3.2	&	\\
ESO059-G001	&	07 31 19.3	&	-68 11 10	&	9	&	4.6$^{c}$	&	TRGB	&	-	&	-	&	17.7$^{a}$	&	1.25	&	14.0$^{f}$	&	1.6	&	\\
ESO115-G021	&	02 37 45.0	&	-61 20 28	&	7	&	5.0$^{f}$	&	TRGB	&	14.6$^{b}$	&	39.80	&	97.6$^{a}$	&	0.09	&	13.3$^{f}$	&	2.5	&	\\
ESO121-G020	&	06 15 54.5	&	-57 43 35	&	10	&	6.1$^{c}$	&	TRGB	&	17.0$^{a}$	&	-	&	14.1$^{a}$	&	0.15	&	15.3$^{g}$	&	1.3	&	\\
ESO154-G023	&	02 56 50.4	&	-54 34 23	&	8	&	5.6$^{g}$	&	TRGB	&	13.8$^{b}$	&	40.36	&	139.2$^{a}$	&	0.06	&	12.8$^{b}$	&	2.4	&	\\
ESO245-G005	&	01 45 03.6	&	-43 35 53	&	9	&	4.4$^{h}$	&	TRGB	&	13.6$^{b}$	&	39.87	&	81.0$^{a}$	&	0.06	&	12.7$^{a}$	&	1.5	&	\\
ESO294-G010	&	00 26 33.3	&	-41 51 20	&	10	&	1.9$^{i}$	&	TRGB	&	18.0$^{b}$	&	38.50	&	0.3$^{c}$	&	0.02	&	15.5$^{f}$	&	3.1	&	\\
ESO308-G022	&	06 39 32.9	&	-40 43 13	&	10	&	7.7	&	h	&	-	&	-	&	3.8$^{a}$	&	0.31	&	16.1$^{e}$	&	2.4	&	\\
ESO347-G017	&	23 26 56.1	&	-37 20 49	&	7	&	7.6$^{e}$	&	TF	&	15.6$^{b}$	&	39.87	&	8.4$^{a}$	&	0.06	&	15.8$^{b}$	&	4	&	\\
ESO348-G009	&	23 49 23.4	&	-37 46 25	&	10	&	11.5$^{e}$	&	TF	&	16.3$^{b}$	&	-	&	13.4$^{a}$	&	0.05	&	14.8$^{i}$	&	2.1	&	\\
ESO349-G031	&	00 08 13.3	&	-34 34 42	&	10	&	3.2$^{c}$	&	TRGB	&	17.0$^{b}$	&	-	&	5.8$^{a}$	&	0.05	&	15.7$^{c}$	&	2.7	&	\\

\noalign{\smallskip} \\ \\ \\ \\ \\ \\ \\ \\ \\ \\ \\ \\ \\ \\ \\ \\ \\ \\ \\ \\ \\ \\ \\ \\

ESO364-G029	&	06 05 45.4	&	-33 04 54	&	9	&	7.6	&	h	&	-	&	-	&	17.2$^{a}$	&	0.16	&	13.8$^{j}$	&	1.7	&	\\
ESO461-G036	&	20 03 57.4	&	-31 40 54	&	10	&	7.8$^{c}$	&	TRGB	&	-	&	-	&	7.5$^{a}$	&	1.05	&	17.1$^{b}$	&	2.3	&	\\
ESO468-G020	&	22 40 43.9	&	-30 47 59	&	-3	&	2.0$^{e}$	&	txt	&	-	&	-	&	-	&	0.05	&	15.9$^{a}$	&	2.7	&	\\
ESO473-G024	&	00 31 22.5	&	-22 45 57	&	10	&	9.9$^{e}$	&	TF	&	16.8$^{b}$	&	-	&	7.2$^{a}$	&	0.07	&	16.4$^{c}$	&	2.6	&	\\
ESO540-G030	&	00 49 21.1	&	-18 04 28	&	10	&	3.4$^{h}$	&	TRGB	&	19.3$^{b}$	&	39.08	&	0.3$^{c}$	&	0.09	&	16.4$^{f}$	&	3.3	&	\\
ESO540-G032	&	00 50 24.6	&	-19 54 25	&	10	&	3.4$^{h}$	&	TRGB	&	19.5$^{b}$	&	38.87	&	0.3$^{c}$	&	0.08	&	16.4$^{f}$	&	3.3	&	\\
ESO565-G003	&	09 23 09.9	&	-20 10 03	&	10	&	7.5	&	h	&	17.5$^{a}$	&	-	&	-	&	0.21	&	15.5$^{b}$	&	2.6	&	\\
IC1574	&	00 43 03.8	&	-22 15 01	&	10	&	4.92$^{h}$	&	TRGB	&	16.6$^{b}$	&	39.40	&	5.4$^{a}$	&	0.06	&	14.9$^{c}$	&	3	&	\\
IC1959	&	03 33 11.8	&	-50 24 38	&	8	&	6.05$^{f}$	&	TRGB	&	14.4$^{b}$	&	40.24	&	27.2$^{a}$	&	0.04	&	13.2$^{a}$	&	2.3	&	\\
IC2038	&	04 08 54.1	&	-55 59 32	&	7	&	19.2$^{e}$	&	TF	&	-	&	-	&	-	&	0.04	&	15.0$^{e}$	&	3.1	&	\\
IC2039	&	04 08 54.1	&	-55 59 32	&	7	&	9.2	&	h	&	20.2$^{a}$	&	-	&	-	&	0.04	&	14.9$^{a}$	&	3.4	&	\\
IC4662	&	17 47 06.3	&	-64 38 25	&	9	&	2.44$^{c}$	&	TRGB	&	12.4$^{b}$	&	-	&	130.0$^{a}$	&	0.25	&	11.8$^{a}$	&	2.9	&	\\
IC5052	&	20 52 06.2	&	-69 12 14	&	7	&	6.03$^{k}$	&	TRGB	&	13.9$^{b}$	&	41.32	&	101.7$^{a}$	&	0.18	&	11.7$^{h}$	&	2.6	&	\\
IC5152	&	22 02 41.9	&	-51 17 43	&	9	&	1.97$^{f}$	&	TRGB	&	12.3$^{b}$	&	40.09	&	97.2$^{a}$	&	0.09	&	11.0$^{b}$	&	2.7	&	\\

IC5332	&	23 34 27.5	&	-36 06 06	&	7	&	7.8$^{e}$	&	mem	&	12.5$^{b}$	&	41.48	&	159.2$^{a}$	&	0.06	&	11.0$^{b}$	&	2.8	&	\\
KK98-73	&	09 12 29.3	&	-24 15 28.0	&	10	&	9.8$^{e}$	&	mem	&	-	&	-	&	-	&	0.66	&	16.4$^{e}$	&	2.8	&	\\
KKS2000-09	&	06 46 56.6	&	-17 56 27	&	?	&	10$^{l}$	&	mem	&	-	&	-	&	31.4$^{a}$	&	1.47	&	16.4$^{k}$	&	3.7	&	\\
KKS2000-55	&	05 50 17.7	&	-10 17 51	&	?	&	10.1	&	h	&	-	&	-	&	54.3$^{a}$	&	2.83	&	-	&	-	&	\\
NGC1311	&	03 20 07.4	&	-52 11 06	&	8	&	5.2$^{g}$	&	TRGB	&	14.9$^{b}$	&	40.07	&	14.6$^{a}$	&	0.08	&	13.3$^{b}$	&	3	&	\\
NGC1313	&	03 18 15.4	&	-66 29 51	&	7	&	4.07$^{n}$	&	TRGB	&	10.6$^{b}$	&	41.84	&	462.7$^{a}$	&	0.38	&	9.7$^{h}$	&	2.6	&	\\
NGC1705	&	04 54 13.7	&	-53 21 41	&	9	&	5.11$^{o}$	&	TRGB	&	13.3$^{b}$	&	40.32	&	15.4$^{a}$	&	0.03	&	12.8$^{a}$	&	2.6	&	\\
NGC1744	&	04 59 58.2	&	-26 01 36	&	7	&	10$^{e}$	&	TF	&	13.0$^{b}$	&	41.22	&	144.3$^{a}$	&	0.15	&	11.6$^{a}$	&	2.2	&	\\
NGC2784	&	09 12 19.4	&	-24 10 18	&	-2	&	9.82$^{p}$	&	SBF	&	14.9$^{b}$	&	-	&	-	&	0.74	&	11.3$^{a}$	&	4.5	&	\\
NGC2835	&	09 17 52.9	&	-22 21 19	&	5	&	10.3$^{e}$	&	TF	&	12.4$^{b}$	&	-	&	127.8$^{a}$	&	0.35	&	11.0$^{a}$	&	3.6	&	\\
NGC2915	&	09 26 11.5	&	-76 37 35	&	9	&	3.78$^{b}$	&	TRGB	&	13.4$^{b}$	&	-	&	108.4$^{a}$	&	0.95	&	13.2$^{e}$	&	2.9	&	\\
NGC3115	&	10 05 14.0	&	-07 43 07	&	-1	&	9.68$^{p}$	&	SBF	&	15.7$^{b}$	&	-	&	-	&	0.17	&	-	&	-	&	\\
NGC7713	&	23 36 15.0	&	-37 56 20	&	6	&	7.8$^{e}$	&	TF	&	-	&	41.41	&	58.6$^{a}$	&	0.06	&	11.5$^{a}$	&	3	&	\\
NGC7793	&	23 57 49.4	&	-32 35 24	&	6	&	3.91$^{h}$	&	TRGB	&	11.1$^{b}$	&	41.68	&	278.5$^{a}$	&	0.07	&	9.7$^{b}$	&	3.2	&	\\
UGCA148	&	09 09 46.6	&	-23 00 33	&	10	&	9.8$^{e}$	&	mem	&	-	&	-	&	-	&	0.59	&	15.6$^{e}$	&	3	&	\\
UGCA15	&	00 49 49.3	&	-21 00 58	&	10	&	3.34$^{h}$	&	TRGB	&	16.8$^{b}$	&	38.90	&	3.9$^{a}$	&	0.06	&	15.2$^{a}$	&	2.4	&	\\
UGCA153	&	09 13 12.1	&	-19 24 31	&	8	&	21.9$^{e}$	&	TF	&	16.3$^{b}$	&	-	&	13.4$^{a}$	&	0.31	&	15.4$^{h}$	&	2.5	&	\\
UGCA162	&	09 21 27.5	&	-22 30 02	&	8	&	18.6$^{e}$	&	TF	&	16.1$^{b}$	&	-	&	32.9$^{a}$	&	0.24	&	14.9$^{b}$	&	2.2	&	\\
UGCA438	&	23 26 27.5	&	-32 23 26	&	10	&	2.18$^{a}$	&	TRGB	&	15.3$^{b}$	&	-	&	-	&	0.05	&	13.9$^{b}$	&	2.7	&	\\
UGCA442	&	23 43 46.0	&	-31 57 33	&	8	&	4.27$^{h}$	&	TRGB	&	14.8$^{b}$	&	39.44	&	50.1$^{a}$	&	0.06	&	13.6$^{a}$	&	2.3	&	\\
\hline																								
\noalign{\smallskip}

MVPB12	&		&		&		&		&		&		&		&		&		&		&	$B-Ks$	&	\\
	& 		& 		& 		& 		&		&		&		&		&		&		&	[mag]	&	\\
	\noalign{\smallskip}

\hline																								
Cam B	&	04 53 06.9	&	+67 05 57	&	10	&	3.2	&	-	&	-	&	-	&	-	&	0.79	&	16.71	&	2.61	&	\\
CGCG 087-33	&	07 42 31.2	&	+16 33 40	&	10	&	7.9	&	-	&	-	&	-	&	-	&	0.12	&	15.34	&	2.85	&	\\
DDO 006	&	00 49 49.3	&	$-$21 00 58	&	10	&	3.2	&	-	&	16.81$^{b}$	&	38.87	&	-	&	0.06	&	15.3	&	2.22	&	\\
DDO 047	&	07 41 55.0	&	+16 48 02	&	8	&	8	&	-	&	14.77$^{b}$	&	-	&	-	&	0.12	&	13.62	&	0.07	&	\\
DDO 099	&	11 50 53.0	&	+38 52 50	&	10	&	2.6	&	-	&	14.78$^{b}$	&	38.86	&	-	&	0.09	&	13.7	&	2.02	&	\\
DDO 167	&	13 13 22.8	&	+46 19 11	&	10	&	4	&	-	&	16.31$^{b}$	&	-	&	-	&	0.04	&	15.45	&	1.96	&	\\
DDO 168	&	13 14 28.6	&	+45 55 10	&	10	&	4.2	&	-	&	14.53$^{b}$	&	39.65	&	-	&	0.05	&	12.97	&	1.84	&	\\
DDO 181	&	13 39 53.8	&	+40 44 21	&	10	&	3.1	&	-	&	15.57$^{b}$	&	38.69	&	-	&	0.02	&	14.22	&	2.62	&	\\
DDO 187	&	14 15 56.5	&	+23 03 19	&	10	&	2.2	&	-	&	16.21$^{b}$	&	38.52	&	-	&	0.08	&	14.38	&	1.84	&	\\
DDO 190	&	14 24 43.5	&	+44 31 33	&	9	&	2.7	&	-	&	14.8$^{b}$	&	39.40	&	-	&	0.04	&	13.1	&	2.43	&	\\
DDO 226	&	00 43 03.8	&	-22 15 01	&	10	&	4.7	&	-	&	16.56$^{b}$	&	39.37	&	-	&	0.05	&	14.9	&	2.95	&	\\
ESO 059-01	&	07 31 19.3	&	-68 11 10	&	9	&	4.4	&	-	&	-	&	-	&	-	&	0.53	&	13.98	&	2.09	&	\\
ESO 121-20	&	06 15 54.5	&	-57 43 35	&	10	&	5.9	&	-	&	-	&	-	&	-	&	0.15	&	15.27	&	2.49	&	\\
ESO 137-18	&	16 20 59.3	&	-60 29 15	&	6	&	6.2	&	-	&	-	&	-	&	-	&	0.89	&	12.23	&	1.46	&	\\
ESO 215-09	&	10 57 30.2	&	-48 10 44	&	10	&	5.1	&	-	&	-	&	-	&	-	&	0.8	&	16.03	&	2.79	&	\\
ESO 223-09	&	15 01 08.5	&	-48 17 33	&	10	&	6.3	&	-	&	-	&	-	&	-	&	0.94	&	13.82	&	3.32	&	\\
ESO 269-58	&	13 10 32.9	&	-46 59 27	&	9	&	3.6	&	-	&	-	&	-	&	-	&	0.4	&	13.29	&	3.54	&	\\
ESO 320-14	&	11 37 53.4	&	-39 13 14	&	10	&	5.9	&	-	&	-	&	-	&	-	&	0.52	&	15.85	&	2.29	&	\\
ESO 321-14	&	12 13 49.6	&	-38 13 53	&	10	&	3.2	&	-	&	16.41$^{b}$	&	37.81	&	-	&	0.34	&	15.21	&	2.51	&	\\
ESO 324-24	&	13 27 37.4	&	-41 28 50	&	8	&	3.6	&	-	&	14.45$^{b}$	&	-	&	-	&	0.41	&	12.91	&	2.21	&	\\
ESO 325-11	&	13 45 00.8	&	-41 51 32	&	10	&	3.3	&	-	&	-	&	-	&	-	&	0.32	&	14.02	&	2.6	&	\\
ESO 349-31	&	00 08 13.3	&	-34 34 42	&	10	&	3.1	&	-	&	17.05$^{b}$	&	-	&	-	&	0.04	&	15.71	&	2.78	&	\\
ESO 379-07	&	11 54 43.0	&	-33 33 29	&	10	&	5	&	-	&	-	&	-	&	-	&	0.27	&	16.6	&	2.88	&	\\
ESO 381-18	&	12 44 42.7	&	-35 58 00	&	10	&	5.1	&	-	&	-	&	-	&	-	&	0.23	&	15.72	&	2.09	&	\\
ESO 381-20	&	12 46 00.4	&	-33 50 17	&	10	&	5.3	&	-	&	15.12$^{b}$	&	-	&	-	&	0.24	&	14.44	&	2.25	&	\\
ESO 384-16	&	13 57 01.6	&	-35 20 02	&	9	&	4.3	&	-	&	18.53$^{b}$	&	-	&	-	&	0.27	&	15.11	&	2.84	&	\\
ESO 444-78	&	13 36 30.8	&	-29 14 11	&	10	&	5.1	&	-	&	-	&	38.98	&	-	&	0.19	&	15.49	&	2.88	&	\\
ESO 444-84	&	13 37 20.2	&	-28 02 46	&	10	&	4.5	&	-	&	16.02$^{b}$	&	-	&	-	&	0.25	&	15.06	&	2.41	&	\\
ESO 461-36	&	20 03 57.4	&	-31 40 54	&	10	&	7.8	&	-	&	-	&	-	&	-	&	1.1	&	17.06	&	2.72	&	\\
GR 8	&	12 58 40.4	&	+14 13 03	&	10	&	1.9	&	-	&	15.21$^{b}$	&	38.38	&	-	&	0.09	&	14.79	&	1.53	&	\\
Ho II	&	08 19 04.0	&	+70 42 51	&	9	&	3.3	&	-	&	12.32$^{b}$	&	40.47	&	-	&	0.12	&	11.1	&	3.51	&	\\
IC 3104	&	12 18 46.1	&	-79 43 34	&	9	&	2.2	&	-	&	-	&	-	&	-	&	1.49	&	13.65	&	3.06	&	\\
IC 4247	&	13 26 44.4	&	-30 21 45	&	10	&	4.8	&	-	&	15.96$^{b}$	&	39.16	&	-	&	0.24	&	14.41	&	2.76	&	\\
IC 4316	&	13 40 18.1	&	-28 53 40	&	10	&	4	&	-	&	16.1$^{b}$	&	-	&	-	&	0.2	&	14.56	&	3.04	&	\\
IC 4662	&	17 47 06.3	&	-64 38 25	&	9	&	2.3	&	-	&	12.43$^{b}$	&	-	&	-	&	0.25	&	11.74	&	2.91	&	\\
IC 5152	&	22 02 41.9	&	-51 17 43	&	9	&	1.9	&	-	&	12.33$^{b}$	&	40.04	&	-	&	0.09	&	11.03	&	2.78	&	\\
KK98 17	&	02 00 09.9	&	+28 49 57	&	10	&	4.9	&	-	&	-	&	-	&	-	&	0.2	&	17.2	&	3.14	&	\\
KK98 182	&	13 05 02.9	&	-40 04 58	&	10	&	5.7	&	-	&	-	&	-	&	-	&	0.37	&	16.33	&	2.64	&	\\
KK98 200	&	13 24 36.0	&	-30 58 20	&	10	&	4.2	&	-	&	-	&	-	&	-	&	0.25	&	16.67	&	2.59	&	\\
KK98 230	&	14 07 10.7	&	+35 03 37	&	10	&	2	&	-	&	18.38$^{b}$	&	38.16	&	-	&	0.05	&	17.5	&	3.14	&	\\
KKH 086	&	13 54 33.6	&	+04 14 35	&	10	&	2.6	&	-	&	19.27$^{b}$	&	38.67	&	-	&	0.1	&	16.88	&	3.12	&	\\
KKH 098	&	23 45 34.0	&	+38 43 04	&	10	&	2.4	&	-	&	17.2$^{b}$	&	38.57	&	-	&	0.45	&	16.7	&	2.79	&	\\
Mrk 178	&	11 33 29.1	&	+49 14 17	&	10	&	3.7	&	-	&	15.36$^{b}$	&	39.14	&	-	&	0.07	&	14.44	&	2.52	&	\\
NGC 1311	&	03 20 07.4	&	-52 11 06	&	8	&	5.4	&	-	&	14.84$^{b}$	&	40.10	&	-	&	0.08	&	13.4	&	2.62	&	\\
NGC 1569	&	04 30 49.1	&	+64 50 53	&	8	&	2.6	&	-	&	9.65$^{b}$	&	-	&	-	&	2.52	&	11.79	&	0.81	&	\\
NGC 2915	&	09 26 11.5	&	-76 37 35	&	9	&	3.6	&	-	&	13.31$^{b}$	&	-	&	-	&	1	&	13.2	&	2.31	&	\\
NGC 3077	&	10 03 21.0	&	+68 44 02	&	9	&	3.7	&	-	&	-	&	41.42	&	-	&	0.24	&	10.62	&	2.83	&	\\
NGC 3738	&	11 35 48.6	&	+54 31 22	&	9	&	4.6	&	-	&	13.77$^{b}$	&	40.57	&	-	&	0.04	&	11.87	&	1.93	&	\\
NGC 4163	&	12 12 08.9	&	+36 10 10	&	9	&	2.8	&	-	&	15.34$^{b}$	&	39.15	&	-	&	0.07	&	13.63	&	2.48	&	\\
NGC 4214	&	12 15 38.9	&	+36 19 39	&	8	&	3	&	-	&	11.47$^{b}$	&	41.43	&	-	&	0.08	&	10.24	&	2.09	&	\\
NGC 5408	&	14 03 21.5	&	-41 22 35	&	9	&	4.9	&	-	&	-	&	-	&	-	&	0.25	&	12.2	&	1.39	&	\\
NGC 6822	&	19 44 57.7	&	-14 48 11	&	10	&	0.5	&	-	&	10.08$^{b}$	&	-	&	-	&	0.84	&	9.31	&	1.12	&	\\
Peg DIG	&	23 28 34.1	&	+14 44 48	&	10	&	0.9	&	-	&	15.71$^{b}$	&	38.47	&	-	&	0.25	&	13.21	&	3.75	&	\\
Sex A	&	10 11 00.8	&	-04 41 34	&	10	&	1.4	&	-	&	12.55$^{b}$	&	39.06	&	-	&	0.16	&	11.86	&	1.97	&	\\
Sex B	&	10 00 00.1	&	+05 19 56	&	10	&	1.4	&	-	&	13.68$^{b}$	&	38.80	&	-	&	0.11	&	11.85	&	1.53	&	\\
UGC 0685	&	01 07 22.3	&	+16 41 02	&	9	&	4.7	&	-	&	16.04$^{b}$	&	39.52	&	-	&	0.21	&	14.22	&	2.45	&	\\
UGC 3755	&	07 13 51.8	&	+10 31 19	&	10	&	7.4	&	-	&	15.6$^{b}$	&	-	&	-	&	0.32	&	14.07	&	2.5	&	\\
UGC 4115	&	07 57 01.8	&	+14 23 27	&	10	&	7.7	&	-	&	15.71$^{b}$	&	-	&	-	&	0.1	&	15.23	&	3.21	&	\\
UGC 4483	&	08 37 03.0	&	+69 46 31	&	10	&	3.5	&	-	&	15$^{b}$	&	39.12	&	-	&	0.12	&	14.95	&	2.61	&	\\
UGC 6456	&	11 28 00.6	&	+78 59 29	&	10	&	4.3	&	-	&	15.78$^{b}$	&	-	&	-	&	0.13	&	14.27	&	1.56	&	\\
UGC 7605	&	12 28 39.0	&	+35 43 05	&	10	&	4.3	&	-	&	-	&	38.65	&	-	&	0.05	&	14.76	&	2.56	&	\\
UGC 8508	&	13 30 44.4	&	+54 54 36	&	10	&	2.5	&	-	&	16.64$^{b}$	&	38.76	&	-	&	0.05	&	14.12	&	2.36	&	\\
UGC 8833	&	13 54 48.7	&	+35 50 15	&	10	&	3	&	-	&	15.69$^{b}$	&	38.85	&	-	&	0.04	&	15.3	&	2.99	&	\\
UGCA 092	&	04 32 00.3	&	+63 36 50	&	10	&	3.1	&	-	&	15.31$^{b}$	&	38.89	&	-	&	2.85	&	15.22	&	1.35	&	\\
UGCA 438	&	23 26 27.5	&	-32 23 26	&	10	&	2.2	&	-	&	-	&	38.73	&	-	&	0.05	&	13.89	&	2.99	&	\\
WLM	&		&		&	?	&	0.9	&	-	&	-	&	-	&	-	&	0.14	&	-	&	-	&	\\
								
\noalign{\smallskip}
\hline
\enddata
\end{deluxetable}

\clearpage

\begin{deluxetable}{lcc ccc cc} 
\tablecaption{Table listing the computed quantities for the NIR catalogue galaxies, and is arranged as follows : Column 1 - Name, Column 2 - $L(FUV)_{\text{corr}}$, Column 3 - $\dot{M}_*$, Column 4 - {$M_{\text{\hi}}$}, Column 5 - $R_{26.5,eq}$, Column 6 - $\Sigma_{\HIsub}$, Column 7 - $\Sigma_{\text{SFR}}$.\label{sampled_appendix}}
\tablewidth{0pt}
\tablehead{
\colhead{Galaxy Name} &
\colhead{$L(FUV)_{\text{corr}}$} &
\colhead{$\dot{M}_*$} &
\colhead{$M_{\HIsub}$} &
\colhead{$R_{26,eq}$} &
\colhead{$\Sigma_{\HIsub}$} &
\colhead{$\Sigma_{SFR}$} \\
	&	[erg	s$^{-1}$]	&	[$M_{\odot}$	yr$^{-1}$]	&	[$M_{\odot}$]	&	[pc]	&	$\log[M_{\odot}$	$pc^{-2}]$	&	$\log[M_{\odot}$	$yr^{-1}$	$Kpc^{-2}]$

}

\startdata

\noalign{\smallskip}
YJLK14	\\																							
\noalign{\smallskip}
\hline																								
AM0106-382	&	-	&	-	&	-	&	1222	&	&	\\													
CEN06	&	-	&	-	&	7.6	&	761	&	1.16	&	-	\\											
ESO149-G003	&	41.24	&	-1.90	&	7.8	&	1761	&	0.22	&	-3.44	\\											
ESO199-G007	&	40.48	&	-2.67	&	7.3	&	1476	&	0.07	&	-3.93	\\											
ESO222-G010	&	-	&	-	&	7.7	&	1708	&	0.66	&	-	\\											
ESO223-G009	&	-	&	-	&	9.0	&	3132	&	1.37	&	-	\\											
ESO252-IG001	&	-	&	-	&	8.1	&	1839	&	0.84	&	-	\\											
ESO269-G058	&	40.67	&	-2.48	&	7.3	&	1801	&	0.08	&	-3.66	\\											
ESO272-G025	&	-	&	-	&	7.7	&	1873	&	0.16	&	-	\\											
ESO274-G001	&	-	&	-	&	8.4	&	5538	&	-0.19	&	-	\\											
ESO318-G013	&	41.01	&	-2.14	&	7.9	&	2310	&	0.23	&	-3.84	\\											
ESO320-G014	&	40.32	&	-2.83	&	7.3	&	1407	&	0.34	&	-3.83	\\											
ESO321-G014	&	40.62	&	-2.53	&	7.2	&	938	&	0.49	&	-3.22	\\											
ESO325-G011	&	-	&	-	&	7.9	&	1594	&	0.66	&	-	\\											
ESO376-G016	&	-	&	-	&	8.1	&	1750	&	0.80	&	-	\\											
ESO379-G007	&	-	&	-	&	7.5	&	1190	&	0.61	&	-	\\											
ESO379-G024	&	-	&	-	&	7.3	&	1098	&	0.40	&	-	\\											
ESO410-G005	&	39.42	&	-3.73	&	5.9	&	601	&	-0.36	&	-3.95	\\											
ESO444-G084	&	41.06	&	-2.09	&	8.0	&	1067	&	1.43	&	-2.68	\\											
ESO489-G?056	&	-	&	-	&	7.1	&	798	&	0.69	&	-	\\											
IC4247	&	41.13	&	-2.02	&	-	&	1697	&	-	&	-3.35	\\											
IC4316	&	40.94	&	-2.21	&	7.0	&	1734	&	-0.18	&	-3.39	\\											
NGC0625	&	42.09	&	-1.06	&	8.0	&	3977	&	-0.03	&	-3.14	\\											
NGC2188	&	-	&	-	&	8.6	&	6623	&	-0.14	&	-	\\											
NGC5264	&	41.50	&	-1.65	&	7.8	&	2246	&	0.41	&	-3.03	\\											
NGC5408	&	-	&	-	&	8.5	&	2714	&	0.84	&	-	\\											
UGCA365	&	40.35	&	-2.80	&	6.9	&	1561	&	-0.59	&	-4.28	\\											
\hline																								
\noalign{\smallskip}
KJRD08	\\																							
\noalign{\smallskip}
\hline																								
AM0319-662	&	39.50	&	-3.65	&	-	&	806	&	-	&	-4	\\											
AM0333-611	&	-	&	-	&	8.4	&	0	&	-	&	-	\\											
AM0521-343	&	-	&	-	&	8.2	&	0	&	-	&	-	\\											
AM0737-691	&	40.98	&	-2.17	&	-	&	0	&	-	&	-	\\											
AM1321-304	&	39.96	&	-3.19	&	6.9	&	895	&	0.26	&	-3.84	\\											
Argo	&	-	&	-	&	8.3	&	2346	&	0.76	&	-	\\											
DDO210	&	39.44	&	-3.71	&	6.4	&	559	&	0.15	&	-3.93	\\											
ESO006-G001	&	-	&	-	&	-	&	1816	&	-	&	-	\\											
ESO059-G001	&	-	&	-	&	7.9	&	2419	&	0.52	&	-	\\											
ESO115-G021	&	41.57	&	-1.57	&	8.8	&	3420	&	0.67	&	-3.66	\\											
ESO121-G020	&	40.81	&	-2.34	&	8.1	&	1566	&	1.07	&	-3.35	\\											
ESO154-G023	&	41.99	&	-1.16	&	9.0	&	7163	&	0.16	&	-4.01	\\											
ESO245-G005	&	41.83	&	-1.32	&	8.6	&	3240	&	0.85	&	-3.04	\\											
ESO294-G010	&	39.51	&	-3.64	&	5.5	&	463	&	-0.51	&	-3.62	\\											
ESO308-G022	&	-	&	-	&	7.7	&	1868	&	0.58	&	-	\\											
ESO347-G017	&	41.53	&	-1.62	&	8.1	&	7388	&	-0.63	&	-4.31	\\											
ESO348-G009	&	41.59	&	-1.56	&	8.6	&	5571	&	0.23	&	-3.94	\\											
ESO349-G031	&	40.17	&	-2.98	&	7.1	&	753	&	0.88	&	-3.25	\\											
ESO364-G029	&	-	&	-	&	8.4	&	4419	&	0.18	&	-	\\											
ESO461-G036	&	-	&	-	&	8.0	&	1416	&	1.01	&	-	\\											
ESO468-G020	&	-	&	-	&	-	&	494	&	-	&	-	\\											
ESO473-G024	&	41.27	&	-1.88	&	8.2	&	2096	&	0.78	&	-3.32	\\											
ESO540-G030	&	39.84	&	-3.31	&	5.9	&	583	&	-0.1	&	-3.36	\\											
ESO540-G032	&	39.66	&	-3.49	&	6.0	&	688	&	-0.42	&	-3.88	\\											
ESO565-G003	&	40.86	&	-2.29	&	-	&	1671	&	-	&	-3.36	\\	
\noalign{\smallskip}
\noalign{\smallskip} \\ \\ \\
IC1574	&	40.81	&	-2.34	&	7.5	&	2015	&	-0.02	&	-3.85	\\											
IC1959	&	41.82	&	-1.33	&	8.4	&	3027	&	0.39	&	-3.31	\\											
IC2038	&	-	&	-	&	-	&	6344	&	-	&	-	\\											
IC2039	&	40.36	&	-2.79	&	-	&	2485	&	-	&	-4.17	\\											
IC4662	&	41.92	&	-1.23	&	8.3	&	1756	&	1.09	&	-2.4	\\											
IC5052	&	42.29	&	-0.86	&	8.9	&	9704	&	-0.35	&	-4.16	\\											
IC5152	&	41.70	&	-1.45	&	7.9	&	2074	&	0.63	&	-2.77	\\											
IC5332	&	42.84	&	-0.30	&	9.4	&	3058	&	1.87	&	-1.79	\\											
KK98-73	&	-	&	-	&	-	&	2136	&	-	&	-	\\											
KKS2000-09	&	-	&	-	&	8.9	&	1931	&	1.61	&	&	\\											
KKS2000-55	&	-	&	-	&	9.1	&	0	&	-	&	-	\\											
NGC1311	&	41.54	&	-1.61	&	8.0	&	2974	&	0.1	&	-3.47	\\											
NGC1313	&	43.27	&	0.12	&	9.3	&	16499	&	0.17	&	-2.96	\\											
NGC1705	&	42.06	&	-1.09	&	8.0	&	2451	&	0.6	&	-2.46	\\											
NGC1744	&	42.88	&	-0.27	&	9.5	&	14165	&	0.33	&	-3.47	\\											
NGC2784	&	42.89	&	-0.26	&	-	&	11817	&	-	&	-3.25	\\											
NGC2835	&	43.35	&	0.20	&	9.5	&	21578	&	0.18	&	-3.12	\\											
NGC2915	&	42.48	&	-0.67	&	8.6	&	2126	&	1.19	&	-2.05	\\											
NGC3115	&	42.90	&	-0.25	&	-	&	0	&	-	&	-	\\											
NGC7713	&	-	&	-	&	8.9	&	8022	&	0.27	&	-	\\											
NGC7793	&	42.85	&	-0.30	&	9.0	&	10673	&	0.23	&	-3.08	\\											
UGCA148	&	-	&	-	&	-	&	2558	&	-	&	-	\\											
UGCA15	&	40.37	&	-2.77	&	7.0	&	1575	&	-0.4	&	-4.19	\\											
UGCA153	&	42.32	&	-0.83	&	9.2	&	8640	&	0.41	&	-3.6	\\											
UGCA162	&	42.21	&	-0.94	&	9.4	&	11452	&	0.11	&	-4.25	\\											
UGCA438	&	40.55	&	-2.60	&	-	&	959	&	-	&	-3.16	\\											
UGCA442	&	41.32	&	-1.82	&	8.3	&	3123	&	0.24	&	-3.91	\\											
\hline	
\noalign{\smallskip}
MVPB12	\\																							
\noalign{\smallskip}
\hline																								
Cam	B	&	-	&	-	&	7.1	&	1037	&	0.55	&	-	\\										
CGCG	087-33	&	-	&	-	&	7.6	&	2285	&	0.37	&	-	\\										
DDO	6	&	40.32	&	-2.83	&	6.9	&	1569	&	0.03	&	-3.91	\\										
DDO	47	&	41.92	&	-1.23	&	9.0	&	2762	&	1.59	&	-2.81	\\										
DDO	99	&	40.94	&	-2.21	&	7.9	&	1423	&	1.07	&	-3.21	\\										
DDO	167	&	40.64	&	-2.51	&	7.2	&	1130	&	0.64	&	-3.31	\\										
DDO	168	&	41.43	&	-1.72	&	8.5	&	3411	&	0.93	&	-3.49	\\										
DDO	181	&	40.71	&	-2.44	&	7.4	&	1266	&	0.71	&	-3.34	\\										
DDO	187	&	40.21	&	-2.94	&	7.1	&	978	&	0.65	&	-3.62	\\										
DDO	190	&	40.96	&	-2.19	&	7.7	&	1231	&	1.00	&	-3.07	\\										
DDO	226	&	40.76	&	-2.38	&	7.5	&	3366	&	-0.04	&	-4.14	\\										
ESO	059-01	&	-	&	-	&	7.9	&	1907	&	0.85	&	-	\\										
ESO	121-20	&	-	&	-	&	8.1	&	1518	&	1.20	&	-	\\										
ESO	137-18	&	-	&	-	&	8.5	&	5071	&	0.62	&	-	\\										
ESO	215-09	&	-	&	-	&	8.9	&	1410	&	2.08	&	-	\\										
ESO	223-09	&	-	&	-	&	8.9	&	4541	&	1.13	&	-	\\										
ESO	269-58	&	-	&	-	&	7.3	&	2702	&	-0.01	&	-	\\										
ESO	320-14	&	-	&	-	&	7.3	&	1317	&	0.58	&	-	\\										
ESO	321-14	&	40.63	&	-2.52	&	7.2	&	1712	&	0.21	&	-3.69	\\										
ESO	324-24	&	41.58	&	-1.56	&	8.2	&	2676	&	0.85	&	-3.12	\\										
ESO	325-11	&	-	&	-	&	7.8	&	2656	&	0.47	&	-	\\										
ESO	349-31	&	40.13	&	-3.02	&	6.8	&	880	&	0.40	&	-3.61	\\										
ESO	379-07	&	-	&	-	&	7.5	&	920	&	1.06	&	-	\\										
ESO	381-18	&	-	&	-	&	7.3	&	1005	&	0.81	&	-	\\										
ESO	381-20	&	41.51	&	-1.64	&	8.3	&	3292	&	0.79	&	-3.37	\\										
ESO	384-16	&	40.06	&	-3.09	&	6.8	&	978	&	0.35	&	-3.77	\\										
ESO	444-78	&	-	&	-	&	7.4	&	1711	&	0.42	&	-	\\										
ESO	444-84	&	41.03	&	-2.12	&	8.0	&	1242	&	1.32	&	-3.01	\\										
ESO	461-36	&	-	&	-	&	8.0	&	1823	&	1.01	&	-	\\										
GR	8	&	40.50	&	-2.65	&	6.8	&	611	&	0.77	&	-2.92	\\										
Ho	II	&	42.16	&	-0.99	&	8.8	&	5068	&	0.93	&	-3.10	\\										
IC	3104	&	-	&	-	&	7.1	&	2303	&	-0.14	&	-	\\										
IC	4247	&	41.11	&	-2.04	&	7.3	&	1950	&	0.19	&	-3.32	\\										
IC	4316	&	40.86	&	-2.29	&	6.9	&	1869	&	-0.14	&	-3.53	\\										
IC	4662	&	41.88	&	-1.27	&	8.2	&	1547	&	1.34	&	-2.34	\\										
IC	5152	&	41.65	&	-1.50	&	7.9	&	2031	&	0.79	&	-2.81	\\										
KK98	17	&	-	&	-	&	6.7	&	1052	&	0.19	&	-	\\										
KK98	182	&	-	&	-	&	7.2	&	1227	&	0.53	&	-	\\										
KK98	200	&	-	&	-	&	6.9	&	685	&	0.69	&	-	\\										
KK98	230	&	39.33	&	-3.82	&	6.4	&	190	&	1.32	&	-3.07	\\										
KKH	86	&	39.50	&	-3.65	&	5.9	&	427	&	0.15	&	-3.61	\\										
KKH	98	&	40.18	&	-2.97	&	6.7	&	520	&	0.81	&	-3.10	\\										
Mrk	178	&	41.00	&	-2.15	&	7.0	&	1172	&	0.36	&	-2.98	\\										
NGC	1311	&	41.58	&	-1.57	&	8.0	&	2959	&	0.56	&	-3.21	\\										
NGC	1569	&	44.85	&	1.70	&	8.1	&	2399	&	0.88	&	0.24	\\										
NGC	2915	&	42.48	&	-0.67	&	8.6	&	2100	&	1.50	&	-2.01	\\										
NGC	3077	&	-	&	-	&	8.9	&	3901	&	1.23	&	-	\\										
NGC	3738	&	41.86	&	-1.29	&	8.0	&	2782	&	0.65	&	-2.87	\\										
NGC	4163	&	40.78	&	-2.37	&	7.2	&	1126	&	0.65	&	-3.17	\\										
NGC	4214	&	42.52	&	-0.63	&	8.8	&	3406	&	1.27	&	-2.39	\\										
NGC	5408	&	-	&	-	&	8.6	&	2933	&	1.14	&	-	\\										
NGC	6822	&	41.87	&	-1.28	&	8.1	&	1449	&	1.24	&	-2.30	\\										
Peg	DIG	&	39.80	&	-3.35	&	6.7	&	786	&	0.41	&	-3.84	\\										
Sex	A	&	41.35	&	-1.80	&	7.9	&	1398	&	1.12	&	-2.79	\\										
Sex	B	&	40.86	&	-2.29	&	7.5	&	1210	&	0.88	&	-3.15	\\										
UGC	685	&	41.06	&	-2.09	&	7.8	&	1573	&	0.95	&	-3.18	\\										
UGC	3755	&	41.68	&	-1.46	&	7.9	&	3678	&	0.32	&	-3.29	\\										
UGC	4115	&	41.50	&	-1.65	&	8.5	&	2812	&	1.08	&	-3.24	\\										
UGC	4483	&	41.13	&	-2.02	&	7.6	&	1111	&	1.01	&	-2.81	\\										
UGC	6456	&	41.00	&	-2.15	&	7.7	&	1264	&	0.95	&	-3.05	\\										
UGC	7605	&	-	&	-	&	7.4	&	1459	&	0.57	&	-	\\										
UGC	8508	&	40.17	&	-2.98	&	7.4	&	1065	&	0.87	&	-3.73	\\										
UGC	8833	&	40.67	&	-2.48	&	7.1	&	838	&	0.77	&	-3.02	\\										
UGCA	92	&	42.99	&	-0.16	&	8.4	&	3142	&	0.89	&	-1.85	\\										
UGCA	438	&	-	&	-	&	7.2	&	986	&	0.73	&	-	\\										
WLM	&	-	&	-	&	7.8	&	0	&	-	&	-	\\
\noalign{\smallskip}
\hline
\enddata
\end{deluxetable}	

\clearpage

\bibliographystyle{mn2e}

\bibliography{bibclean}